\documentclass[%
 preprint,
 superscriptaddress,
 nofootinbib,
 amsmath,amssymb, 
 aps,
]{revtex4-1}

\usepackage{graphicx}% Include figure files
\usepackage{dcolumn}% Align table columns on decimal point
\usepackage{verbatim}% multi-line comments
\usepackage{bm, relsize}% bold math
\usepackage{hyperref}% add hypertext capabilities
\graphicspath{{./figures/}}

\newcommand{\vlowk}{V_{{\rm low}\,k}}

\newcommand{\kf}{k_{\text{F}}}

\newcommand{\fpi}{f_{\pi}}

\newcommand{\lchi}{\Lambda_{\chi}}

\newcommand{\Qtp}{\mathcal{Q}_{+}}
\newcommand{\Qth}{\mathcal{Q}_{-}}
\newcommand{\Qav}{\overline Q}
\newcommand{\Qavp}{{\overline Q}_{+}}
\newcommand{\Qavh}{{\overline Q}_{-}}
\newcommand{\Pav}{P_{\rm av}}
\newcommand{\Wav}{W_{\rm av}}
\newcommand{\kav}{k_{\rm av}}

\newcommand{\jav}{j_{\rm av}}

\newcommand{\kernel}{K}

\newcommand{\fm}{\, \text{fm}}
\newcommand{\fmi}{\, \text{fm}^{-1}}

\newcommand{\la}{\langle}
\newcommand{\ra}{\rangle}

\newcommand{\refstr}{| \Phi_0 \ra}
\newcommand{\refstl}{\la \Phi_0 |}

\newcommand{\Qh}{Q_{-}}
\newcommand{\Qp}{Q_{+}}

\newcommand{\C}{\mathcal{C}}
\newcommand{\s}{\sigma}
\newcommand{\is}{\tau}
\newcommand{\vnn}{V_{NN}}
\newcommand{\vnna}{\mathcal{V}_{NN}}
\newcommand{\eft}{\chi\text{EFT}}
\newcommand{\anti}{\mathcal{A}_{123}}
\newcommand{\vnnn}{V_{3N}}
\newcommand{\vnnnd}{\overline{V}_\text{3N}}
\newcommand{\ve}{V_E}
\newcommand{\ved}{\overline{V}_E}
\newcommand{\threecut}{\Lambda_{3N}}
\newcommand{\tr}{\text{Tr}}
\newcommand{\NNLO}{\ensuremath{{\rm N}{}^2{\rm LO}}}
\newcommand{\nb}{\overline{n}}

\newcommand{\noratio}{\mathcal{R}}
\newcommand{\osz}{^1S_0}
\newcommand{\tso}{^3S_1-^3D_1}

\newcommand\ddfrac[2]{\frac{\displaystyle #1}{\displaystyle #2}}

\newcommand{\pmax}{p_{\text{max}}}

\newcommand{\Wvec}{\mathbf{W}}

\newcommand{\kvec}{\mathbf{k}}

\newcommand{\jvec}{\mathbf{j}}

\newcommand{\pvec}{\mathbf{p}}
\newcommand{\Pvec}{\mathbf{P}}

\newcommand{\khat}{\mathbf{\hat{k}}}
\newcommand{\jhat}{\mathbf{\hat{j}}}

\newcommand{\bbar}{\overline{B}}
\newcommand{\bebar}{\overline{\beta}}

\newcommand{\edd}{E_2^{\text{DD}}}

\newcommand{\eres}{E_2^{\text{RE}}}

\newcommand{\hhps}{\Omega}

\newcommand{\beq}{\begin{equation}}
\newcommand{\eeq}{\end{equation}}
\newcommand{\beqn}{\begin{equation}}
\newcommand{\eeqn}{\end{equation}}
\newcommand{\bseq}{\begin{subequations}}
\newcommand{\eseq}{\end{subequations}}

\newcommand{\bi}{\begin{itemize}}
\newcommand{\ei}{\end{itemize}}

\newcommand{\be}{\begin{enumerate}}
\newcommand{\ee}{\end{enumerate}}
\newcommand{\bc}{\begin{center}}
\newcommand{\ec}{\end{center}}

\newcommand\numberthis{\addtocounter{equation}{1}\tag{\theequation}}

\begin{document}

\bibliographystyle{apsrev4-1}

\preprint{APS/123-QED}

\title{Estimates and power counting in uniform 
	matter \texorpdfstring{\\}{} with softened interactions}

\author{A. Dyhdalo}
\email{dyhdalo.2@osu.edu}
\affiliation{%
Department of Physics, The Ohio State University, Columbus, OH 43210
}%

\author{S.K.~Bogner}
\email{bogner@nscl.msu.edu}
\affiliation{% 
National Superconducting Cyclotron Laboratory and Department of 
Physics and Astronomy, Michigan State University, East Lansing, MI 48824, USA}

\author{R.J. Furnstahl}
 \email{furnstahl.1@osu.edu}
\affiliation{%
Department of Physics, The Ohio State University, Columbus, OH 43210
}%

\date{\today}

\begin{abstract}
 Modern softened nucleon-nucleon interactions are well-suited for perturbative many-body calculations, but a many-body power counting scheme is lacking.
 Estimates of diagrammatic contributions at finite density are important ingredients in such a scheme.
 Here we show how to make
 quantitative estimates of the particle-particle and hole-hole channel in uniform nuclear matter for soft interactions. 
 We also use estimates to assess the role of normal-ordered three-body forces for a pure contact interaction.
\end{abstract}

\maketitle

%\tableofcontents

\section{Introduction}
 \label{sec:intro}

Diagrammatic power counting assigns an expansion order to individual Feynman (or other) diagrams
according to their expected relative contribution.
For diagrams at finite density,
such assignments depend critically on the nature of the potential, which in turn leads to
different types of expansion.
The original work on the Brueckner-Bethe-Goldstone (BBG) method for the nuclear many-body problem
included a form of power counting based on estimates of the relative sizes of Goldstone
diagram contributions to the energy per particle in uniform matter.  These estimates motivated 
the hole-line expansion in terms of resummed G matrices~\cite{Day:1967zz,Rajaraman:1967zza,Day:1978,baldo1999nuclear},
but assumed a nucleon-nucleon ($NN$) potential with a strongly repulsive core.
Modern interactions based on chiral effective field theory 
($\eft$)~\cite{Machleidt:2011zz,Epelbaum:2008ga,Gezerlis:2014zia,PhysRevLett.115.122301,PhysRevC.91.024003} 
and/or renormalization group (RG) evolution are much softer and lead to dramatically 
different contributions of individual diagrams, which enables a many-body perturbation theory (MBPT)
expansion.
This difference is also relevant for nonperturbative many-body methods that use basis expansions
(for recent theoretical developments on calculations in uniform matter, see e.g., Refs.~\cite{Bogner:2005sn,
Hebeler:2010xb,
Baldo:2012,
PhysRevC.88.054312,
PhysRevC.89.014319,
PhysRevC.90.054322,
PhysRevC.91.054311,
PhysRevC.93.055802,
PhysRevC.94.054307}). 
In this paper we make progress toward a robust and systematic power counting for softened
interactions in uniform matter by showing how to estimate individual terms in the 
particle-particle (pp) and hole-hole (hh) ladders.
  
In estimating diagrams for uniform matter in MBPT, we emphasize the role of the finite 
density geometric phase space and make approximations such that the momentum 
integrations for a given diagram factorize.
These approximations simplify calculations but yield good quantitative estimates of different 
terms in MBPT and their scaling behavior in the ladder.
Note that we do not require \emph{high precision} values of terms in MBPT, but instead seek to
capture general quantitative behavior so as to motivate a systematic power counting and allow for credible error estimates.

Throughout this work we use the Argonne $v_{18}$ (AV18) interaction~\cite{wiringa:av18} in the $\osz$ and $\tso$ partial waves softened to various degrees with the similarity renormalization group (SRG)~\cite{Bogner:2009bt}. 
AV18 is chosen as a representative hard $NN$ potential for which the power counting in the pp ladder drastically changes under RG transformations.
For coordinate-space potentials such as AV18, hardness is associated with large matrix elements 
at small relative distance, i.e., the repulsive core, and the intermediate-range tensor force. 
In momentum representation, interactions are deemed hard if they strongly couple states of high and low momentum.
The decoupling of these states via the SRG is achieved by a series of unitary transformations characterized by a flow parameter $\lambda$.
Here we make the common choice of the relative kinetic energy in the SRG generator such that 
as $\lambda$ decreases toward zero, the potential flows to band diagonal form~\cite{Bogner:2009bt}.
An alternative would be to use a block-diagonal generator~\cite{Anderson:2008mu}, which reproduces the low-momentum
structure of $\vlowk$ potentials and can be treated with similar estimates.

The evolution to smaller $\lambda$ for different initial $NN$ interactions that are phase equivalent
and share the same long-distance (pion) physics drives the partial wave matrix elements
toward a universal form, up to the momentum scale at which the phase shifts 
agree~\cite{Bogner:2001gq,PhysRevC.89.014001}.
This includes the matrix elements that determine the diagrammatic contributions at least as
high as nuclear matter saturation density.
Thus, even though we use AV18 as the initial potential, our quantitative results for lower
values of $\lambda$ will be the same for other initial potentials such as those based on $\eft$,
and so our conclusions should be quite general.
  
Previous work has established how the nonperturbative nature of $NN$ interactions is
modified by the softening with $\lambda$ combined with the effects of finite density~\cite{Bogner:2005sn,Bogner:2009bt,Hebeler:2010xb,Furnstahl:2013oba}.
Forces such as AV18 are nonperturbative in free space for several reasons: a strong short-range 
repulsive core, iterated tensor components, and the fine-tuning that
produces weakly bound or just unbound states.
The latter requires some form of nonperturbative resummation independent of the details 
of the potential (see e.g.,~\cite{Schafer:2005kg}).
For potentials with nonperturbative repulsive cores, Pauli blocking in uniform matter
does not change the need for resummation because the repulsive cores ensure that 
contributions well above the Fermi surface dominate. 
This meant that the BBG method for uniform matter started with the sum of pp ladder
diagrams to all orders.
On the other hand, Pauli blocking is effective in de-tuning the bound or near-bound
states at densities well below nuclear saturation density. 

One method of explicitly verifying when resummation is needed  
and assessing perturbativeness in general is the Weinberg eigenvalue approach~\cite{Weinberg:1963zz}, 
which in free space examines eigenvalues of the Born series for the Lippmann-Schwinger equation 
and has been extended to finite density.
The analysis of Weinberg eigenvalues in uniform systems has indicated that softened 
interactions become perturbative with increasing density, at least in
the particle-particle channel~\cite{Bogner:2005sn,Hebeler:2010xb}. 
A related set of eigenvalues at finite density arises in our estimation method
and provides similar diagnostics, with a direct connection to the
evaluation of diagrams.
We note that MBPT convergence of soft interactions has been demonstrated in finite
nuclei when using a Hartree-Fock reference state~\cite{Tichai2016283} and in uniform
neutron matter using various nonperturbative 
many-body methods~\cite{Gezerlis:2014zia,PhysRevLett.111.032501,PhysRevC.88.054312,
PhysRevC.89.014319,PhysRevC.90.054322}.
For symmetric nuclear matter including three-body forces, MBPT is somewhat less
perturbative~\cite{PhysRevC.89.014319}, but further investigation is needed.

	Although the nuclear matter power counting analysis of BBG was focused on $NN$ interactions, $3N$ forces have been established as playing an essential role in nuclear matter saturation with modern potentials~\cite{Hebeler:2010xb,
	Bogner:2005sn,
	PhysRevC.91.051301}. 
	Hence, assessing the contributions of $3N$ forces is a crucial task in creating a consistent and systematic many-body power counting. 
	For simplicity, here we limit ourselves to estimates for the size of normal-ordered three-body (effective two-body) contributions compared to residual three-body terms using a pure three-body contact as shows up at \NNLO~in $\eft$.

	The paper is organized as follows: In Sec.~\ref{sec:mbpt} we examine and estimate various diagrams for two-body interactions.  
	In Sec.~\ref{sec:unitary} we briefly show why the conclusions in Sec.~\ref{sec:mbpt} do not apply to the unitary gas.
	In Sec.~\ref{sec:three_body} we discuss three-body forces and give estimates for normal-ordered and residual terms.
	Our findings are summarized in Sec.~\ref{sec:conclusion}.
	Diagrammatic rules and useful formulas are given in the Appendices.

\section{Softened \texorpdfstring{$NN$}{NN} Interactions}
 \label{sec:mbpt}

	In this section we discuss estimates for $NN$ interactions in uniform matter, with estimates for $3N$ forces considered in Sec.~\ref{sec:three_body}.
	We first briefly review MBPT (see e.g., Refs.~\cite{baldo1999nuclear,Shavitt:2009,FETTER71}), discuss different quantities appearing in Goldstone diagrams, and then apply our averaging techniques to the pp channel. 
 
\subsection{Review of MBPT}

\begin{figure}[t]
 \begin{center}
  \includegraphics[width=3.0in]{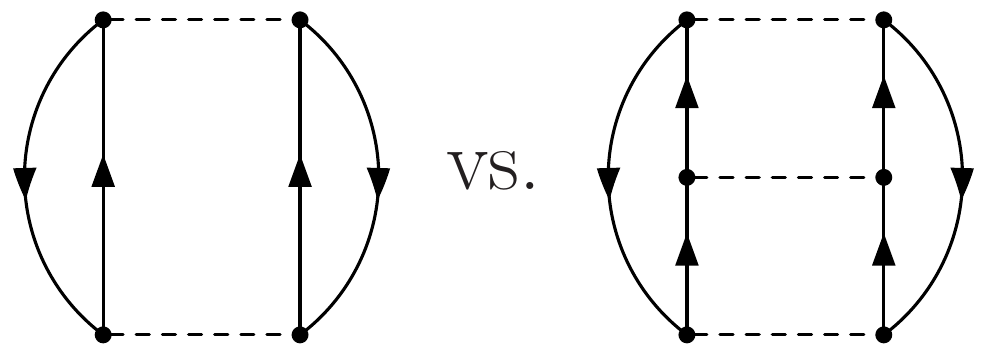}
 \end{center}
 \caption{Second- and third-order Goldstone diagrams, for which we consider their relative size. 
Particles are upward-going arrows, holes are downward-going arrows, and dashed lines are two-body potential insertions.}
 \label{fig:second_third_order}
\end{figure}

	When performing perturbation theory for a given Hamiltonian $H$, one splits $H$ into two parts: an exactly solvable part $H_0$ and a remaining piece $H_I$ such that,
\beq
	H = H_0 + H_I \; ,
	\qquad
	H_0 \refstr = E_0 \refstr \; ,
\eeq
where $H_0$ defines a reference state $\refstr$.
	For our purposes, we identify $H_I$ as the $NN$ potential and adopt a spin-saturated, isospin-symmetric reference state of non-interacting fermions filled up to Fermi momentum $\kf$, 
\beq
	H_I = \vnn
	\; ,
	\qquad
	\refstr = \prod_{i=1}^A
	a^{\dagger}_i | 0 \ra
	\; ,
	\label{eq:ref_state}
\eeq
where the $a^{\dagger}$ operators obey anti-commutation relations and the lowest $A$ orbitals in the Fermi sea are filled.
	Although commonly used to speed convergence in nuclear matter calculations with hard interactions, 
	in this work we have not included a one-body potential in our $H_0$ and $H_I$ terms.  
	The linked cluster expansion~\cite{Goldstone:1957}, allows for the energy of an interacting system to be expressed as an expansion around the reference state,
\beq
	E = E_0 + \refstl H_I \sum_{n=0}^{\infty} 
	\left( \frac{1}{E_0 - H_0} H_I \right)^n 
	\refstr_{\text{connected}} \; ,
\label{eq:goldstone_theorem}
\eeq
	where $H_I$ in 
Eq.~\eqref{eq:goldstone_theorem} is now understood to create particles and holes with respect to the reference state.
	Expressions for individual contributions in the series of Eq.~\eqref{eq:goldstone_theorem} have a simple diagrammatic representation in Goldstone diagrams.%
\footnote{We employ antisymmetrized Goldstone diagrams throughout, where each dashed line represents an antisymmetrized matrix element, see Ref.~\cite{Shavitt:2009}.}
	The subscript `connected' in Eq.~\eqref{eq:goldstone_theorem} ensures that the reference state $\refstr$ does not contribute as an intermediate state, and means that disconnected diagrams do not contribute to the energy. For our purposes, we want to consider the \textit{relative} importance of different diagrams, for example the relative sizes of the two diagrams in Fig.~\ref{fig:second_third_order}.
	A list of rules for translating Goldstone diagrams into mathematical expressions is given in Appendix \ref{sec:rules}.

	Goldstone diagrams differ from Feynman diagrams in that they are time-ordered, thus each Feynman diagram corresponds to multiple Goldstone diagrams. 
	Our use of Goldstone diagrams is historically  motivated by their original use in BBG theory due to the asymmetry in the power counting of particle and hole lines. 
	We do not assess here whether Feynman diagrams might be a more efficient approach for sufficiently softened interactions.

\subsection{Averaging and Approximations}

	In this section, we define relevant quantities appearing in Goldstone diagrams as well as their averaged counterparts. 
	Pauli blocking operators for intermediate particle and hole states are defined as, respectively, 
\bseq
\beq
	\Qp (\Pvec/2, \kvec; \kf) 
	\equiv 
	\nb (\Pvec/2 + \kvec)
	\;
	\nb (\Pvec/2 - \kvec) \; ,
\eeq
\beq
	\Qh (\Pvec/2, \kvec; \kf) 
	\equiv 
	n (\Pvec/2 + \kvec)
	\;
	n (\Pvec/2 - \kvec) \; ,
\eeq
\eseq
using the distribution functions defined in Appendix \ref{sec:rules}.
	Pauli blockers can be angle-averaged,
\beq
	\Qav_{\pm} (P, k ; \kf)
	=
	\frac{1}{4\pi}
	\int
	d\Omega_{\kvec} 
	\;
	Q_{\pm} (\Pvec/2, \kvec; \kf) \; ,
\eeq
and are then given by,
\bseq
\beq
	\Qavp =
	\begin{cases}
	0 & \text{for } k < \sqrt{\kf^2-P^2/4} \\
	1 & \text{for } k > \kf + P/2 \\
	A &
	\text{otherwise} 
	\end{cases}
	\; ,
	\qquad
	\Qavh = 
	\begin{cases}
	0 & \text{for $k > \sqrt{\kf^2-P^2/4}$} \\
	1 & \text{for $k < \kf - P/2$} \\
	- A &
	\text{otherwise} 
	\end{cases}
	\; ,
	\label{eq:NN_angleavg}
\eeq
where
\beq
	A \equiv  
	\frac{\textstyle k^2 + P^2/4 - \kf^2}
	{\textstyle kP} \; . 
\eeq
\eseq
	For potentials with no angular dependence (s-wave), the above procedure is exact.
	We also make use of the hole phase space found after integrating over the total momentum $P$,
\beq
	\hhps \left(\frac{k}{\kf}, \kf \right)
	\equiv
	\int dP \; P^2 \; \Qavh (P, k ; \kf)
	\; ,
	\qquad
	\hhps \left(x, \kf \right)
	=
	\frac{4 \kf^3}{3}
	\left(
	2 - 3 x + x^3
	\right) 
	\Theta(1 - x) \; .
\eeq

	The mean square average of a quantity in our system, say the total two-body momentum, is defined in the usual way, 
\beq
	\la P^2 \ra = 
	\ddfrac{\int d^3\pvec_1 \; d^3\pvec_2 
	\; \left(\pvec_1 + \pvec_2 \right)^2 \; 
	n(\pvec_1) n(\pvec_2) }
	{\int d^3\pvec_1 \; d^3\pvec_2 \; 
	n(\pvec_1) n(\pvec_2)} \; .
	\label{eq:rms_proc}
\eeq
	This results in the root mean square (RMS) total momentum $\Pav$ and RMS hole relative momentum $\kav$,
\beq
	\Pav = \sqrt{\frac{6}{5}} \kf \; ,
	\qquad
	\kav = \sqrt{\frac{3}{10}} \kf
	\; .
	\label{eq:rms_nn}
\eeq

\subsection{Particle-Particle Channel}

	In this section, we explicitly calculate different terms in the pp ladder and show how to extract quantitative estimates.
	In the following, we restrict ourselves to s-wave channels as their net contribution dominates the energy density of nuclear matter over the net contribution of other channels. 
	The energy per particle of the $n$th rung in the pp ladder, excluding $n=1$ (Hartree-Fock), is given by, 
\begin{align*}
	\frac{E^{(n)}_{\rm pp}}{N}
	&=
	\left(
	\frac{1}{2}
	\right)^{n}
	\left(\frac{2}{\pi}\right)^{n}
	2^{n}
	\left(\frac{m}{\hbar^2} \right)^{n-1}
	\frac{
	\left(
	-1
	\right)^{n-1}
	}{\rho}
	\int 
	\frac{d^3\Pvec}{(2\pi)^3}
	\int dk_1 \; k_1^2 \cdots
	\int dk_{n} \; k_{n}^2 \;
	(2T + 1)
	(2J + 1)
	\\
	&\times
	\ddfrac{\Qavh (P, k_1; \kf) \;
	\Qavp (P, k_2; \kf)
	\cdots
	\Qavp (P, k_{n}; \kf)
	}
	{(k_2^2 - k_1^2) 
	\cdots 
	(k_n^2 - k_1^2)}
	\enspace
	\la k_1 | V | k_2 \ra 
	\cdots
	\la k_n | V | k_1 \ra
	\; ,
	\numberthis
	\label{eq:pp_ladder_energy}
\end{align*}
	where the $1/2$ are symmetry factors, the $2/\pi$ comes from the partial wave basis expansion, the $2$ from antisymmetry of the potential, the $m/\hbar^2$ from the energy denominators, the $(-1)$ from flipping the energy denominator arguments, the $(2T+1)$ and $(2J+1)$ from the $T_z$ and $J_z$ sums, all Pauli operators are angle-averaged, and $\la k_a | V | k_b \ra$ are momentum space potential matrix elements in a given partial wave including coupled channels (see Appendix \ref{sec:pw_appendix} for details).
	Also note that in Eq.~\eqref{eq:pp_ladder_energy} we have chosen a free single-particle energy spectrum. 

	We assume that the energy integrand in Eq.~\eqref{eq:pp_ladder_energy} is dominated by phase-space
	regions where the particle relative momentum $k'$ is sufficiently larger than the total momentum $P$
    and the hole relative momentum $k$ such that $k'$ will primarily drive the behavior of energy denominators and particle Pauli blockers. This motivates the approximations
\beq
	\frac{1}{k'^2 - k^2} \approx
	\frac{1}{k'^2 - \kav^2}
	\; , \qquad
	\Qavp (P, k' ; \kf) \approx
	\Qavp (\Pav, k' ; \kf)
	\; ,
	\label{eq:nn_approx}
\eeq
where we have used the RMS values in Eq.~\eqref{eq:rms_nn}. 
	To facilitate calculations, we also render the expression in Eq.~\ref{eq:pp_ladder_energy} on a discrete mesh for the momentum integrations,
\begin{align*}
	\frac{E^{(n)}_{\rm pp}}{N}
	&=
	\frac{2}{\pi}
	\left(
	\frac{2 m}{\pi \hbar^2}
	\right)^{n-1}
	\frac{	
	\left(
	-1
	\right)^{n-1}
	}{2 \pi^2 \rho} \;
	\sum_{k_i}
	k_1^2 \ w_1 \cdots
	k_{n}^2 \ w_n \;
	(2T + 1)
	(2J + 1) \
	\hhps \left(\frac{k_1}{\kf}, \kf \right)
	\\
	&\times
	\ddfrac{\Qavh (P, k_1; \kf) \;
	\Qavp (P, k_2; \kf)
	\cdots
	\Qavp (P, k_{n}; \kf)
	}
	{(k_2^2 - k_1^2) 
	\cdots 
	(k_n^2 - k_1^2)}
	\enspace
	\la k_1 | V | k_2 \ra 
	\cdots
	\la k_n | V | k_1 \ra
	\; ,
	\numberthis
	\label{eq:pp_energy_discrete}
\end{align*}
where $w_i$ refers to the relevant weight for a momentum sum. 

	The two approximations in Eq.~\eqref{eq:nn_approx} allow for the momentum integrations in our ladder to factorize, connected only by potential matrix elements.
	Reorganizing Eq.~\eqref{eq:pp_energy_discrete} yields,

\bseq
\begin{align*}
	\frac{E^{(n)}_{\rm pp}}{N}
	&\approx
	\frac{2}{\pi}
	\left(
	\frac{2 m}{\pi \hbar^2}
	\right)^{n-1}
	\frac{	
	\left(
	-1
	\right)^{n-1}
	}{2 \pi^2 \rho} \;
	\sum_{k_i} \;
	(2T + 1)
	(2J + 1) \;
	\hhps \left(\frac{k_1}{\kf}, \kf \right) \;
	k_1^2 \ w_1
	\\
	&\times
	\la k_1 | V | k_2 \ra
	\sqrt{
	\frac{ \; 
	\Qavp (\Pav, k_2; \kf) \; k_2^2 \ w_2}
	{k_2^2 - \kav^2}	
	}
	\;
	\kernel^{n-2}
	\;
	\sqrt{
	\frac{
	\Qavp (\Pav, k_n; \kf) \; k_n^2 \ w_n \; 
	}
	{k_n^2 - \kav^2}	
	}
	\la k_n | V | k_1 \ra \; ,
	\numberthis
	\label{eq:pp_est}
\end{align*}
where $K$ is a kernel for particle-particle scattering,

\beq
	\kernel \equiv 	
	\sqrt{
	\frac{\Qavp (\Pav, k_a; \kf) \; k_a^2 \ w_a}
	{k_a^2 - \kav^2}	
	}
	\la k_a | V | k_b \ra
	\sqrt{
	\frac{\Qavp (\Pav, k_b; \kf) \; k_b^2 \ w_b}
	{k_b^2 - \kav^2}	
	} \; .
	\label{eq:pp_kernel}
\eeq
\eseq
	Factorization via our two approximations ensures that adding more rungs to the ladder corresponds to additional powers of the pp kernel $\kernel$ without affecting the outer parts of the integrand in Eq.~\eqref{eq:pp_est}. 
	The particle phase space for the interior parts of the ladder has thus been completely decoupled from the hole phase space.

	The kernel $\kernel $ in Eq.~\eqref{eq:pp_kernel} is real and symmetric like the potential and so can be diagonalized in an eigendecomposition,
\beq
	\kernel^n = L D^n L^{-1} \; ,
	\label{eq:kernel_diag}
\eeq
where $D$ is a diagonal matrix holding the kernel eigenvalues and $L$ is a matrix of the kernel eigenvectors.
	Because $D$ is a diagonal matrix, in this decomposition successive rungs of the ladder correspond to simple powers of the kernel eigenvalues.
	As such, Eq.~\eqref{eq:kernel_diag} allows for high orders in the ladder to be computed with little additional computational cost.
	
\begin{figure}[t]
  \includegraphics[width=1.0\textwidth]{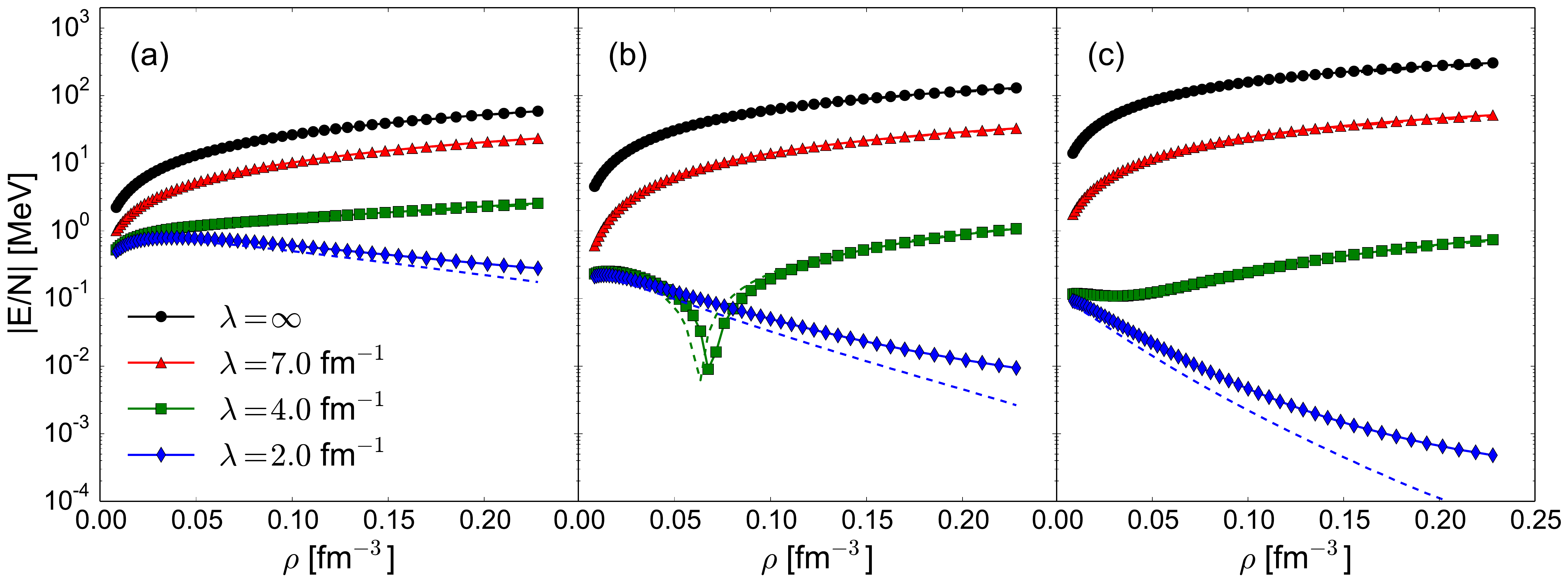}~
  \caption{\textbf{(a)} The absolute value of the second-order energy per particle in nuclear matter is plotted as a function of density $\rho$ for the $\osz$ partial wave using the AV18 potential.  
  Both exact (solid) and estimates (dashed) are shown for four different SRG $\lambda$ scales.
 \textbf{(b)} Same as (a) but for third-order in the pp channel.
 \textbf{(c)} Same as (a) but for fourth-order in the pp channel.}
 \label{fig:pp_1S0_AV18}
\end{figure}

\begin{figure}[t]
  \includegraphics[width=1.0\textwidth]{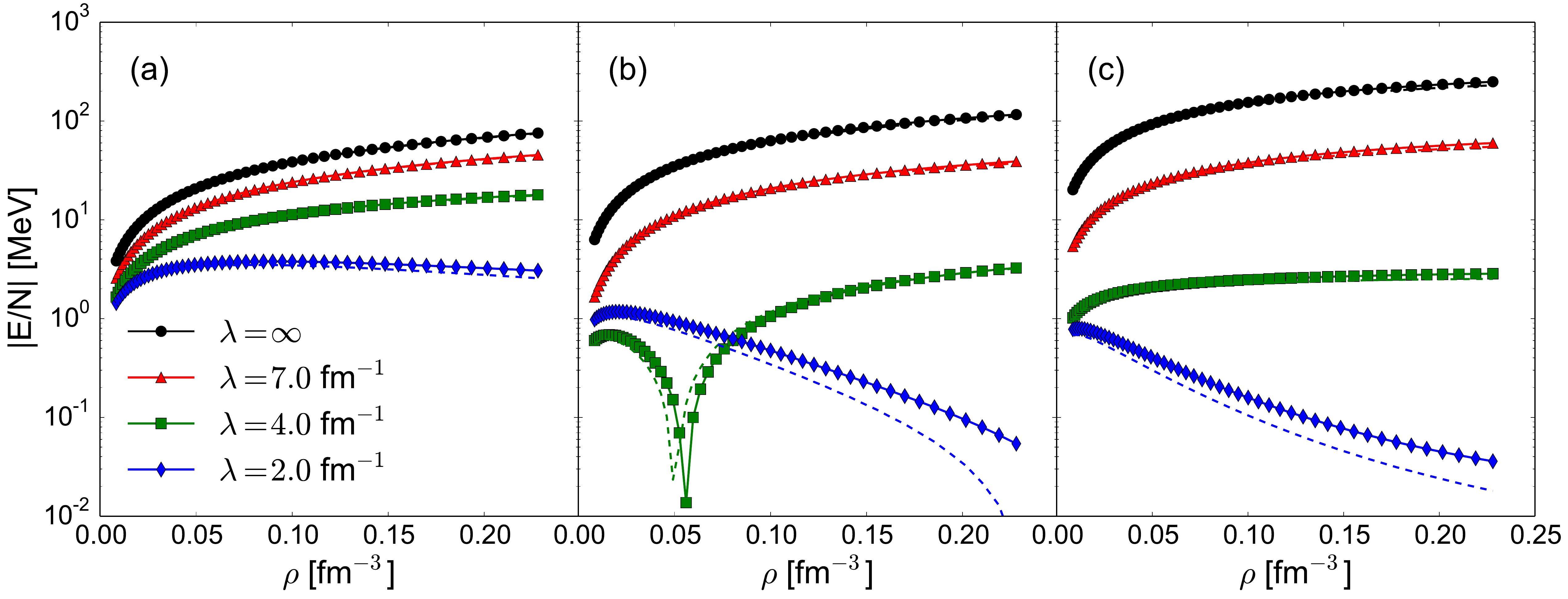}~
  \caption{\textbf{(a)} The absolute value of the second-order energy per particle in nuclear matter is plotted as a function of density $\rho$ for the $\tso$ partial wave using the AV18 potential.  
  Both exact (solid) and estimates (dashed) are shown for four different SRG $\lambda$ scales.
 \textbf{(b)} Same as (a) but for third-order in the pp channel.
 \textbf{(c)} Same as (a) but for fourth-order in the pp channel.}
 \label{fig:pp_3S1_AV18}
\end{figure}

	In Fig.~\ref{fig:pp_1S0_AV18} we show the absolute value of the second-order energy per particle and third- and fourth-order terms in the pp ladder for nuclear matter for the $\osz$ partial wave using the AV18 \footnote{When using AV18 we assume full isospin symmetry and use the np force.} potential.
	The energy terms are calculated for the potential evolved to four different SRG $\lambda$ scales. 
	Here $\lambda=\infty$ refers to the unevolved AV18 potential whereas the evolution proceeds further for lower $\lambda$.
	Fig.~\ref{fig:pp_3S1_AV18} then shows the same quantities in the pp ladder but for the $\tso$ partial wave using the AV18 potential.
	Both exact calculations from Eq.~\eqref{eq:pp_energy_discrete} and estimates using Eq.~\eqref{eq:pp_est} are shown. 
	In keeping with previous results~\cite{Hebeler:2010xb,Bogner:2005sn}, the trend is for the relative importance of higher orders in MBPT to decreases as $\lambda$ lowers.
	In all six plots, our estimates do a good job of reproducing the exact results, suggesting our approximations are well motivated and capture the relevant physics.
	In Table \ref{tab:absolute_errors}, we list the absolute errors induced by our approximations for $\lambda = 4.0$ and $2.0 \fmi$ near saturation density $\rho \approx 0.16 \; \text{fm}^{-3}$.
	To avoid repetition, we give calculations, estimates, and discussion of the hh ladder in Appendix~\ref{sec:hole_hole}.

\begingroup
%\squeezetable
\begin{table}[t]
 \caption{\label{tab:absolute_errors}
 List of the absolute differences between the exact and estimate calculations for the energy per particle of diagrams in the pp ladder for AV18. All quantities below are in MeV, are evaluated near the saturation point $\rho = 0.163 \fm^{-3}$, and are rounded to the nearest decimal. 
}
 \begin{ruledtabular}
 \begin{tabular}{c ||  c | c | c }
  &  Second-Order &  Third-Order & Fourth-Order  \\
 \hline
 $\osz \quad \lambda = 4.0 \fmi$ & 0.088 & 0.031 & 6.8E-3  \\
 $\osz  \quad \lambda = 2.0 \fmi$ & 0.11  & 0.010 & 8.6E-4 \\
 \hline
 $\tso \quad \lambda = 4.0 \fmi$ & 0.17 & 0.048 & 0.27  \\
 $\tso \quad \lambda = 2.0 \fmi$ & 0.43 & 0.085 & 0.028 \\
 \end{tabular}
 \end{ruledtabular}
\end{table}
\endgroup

\begin{figure}[t]
  \includegraphics[width=0.49\textwidth]{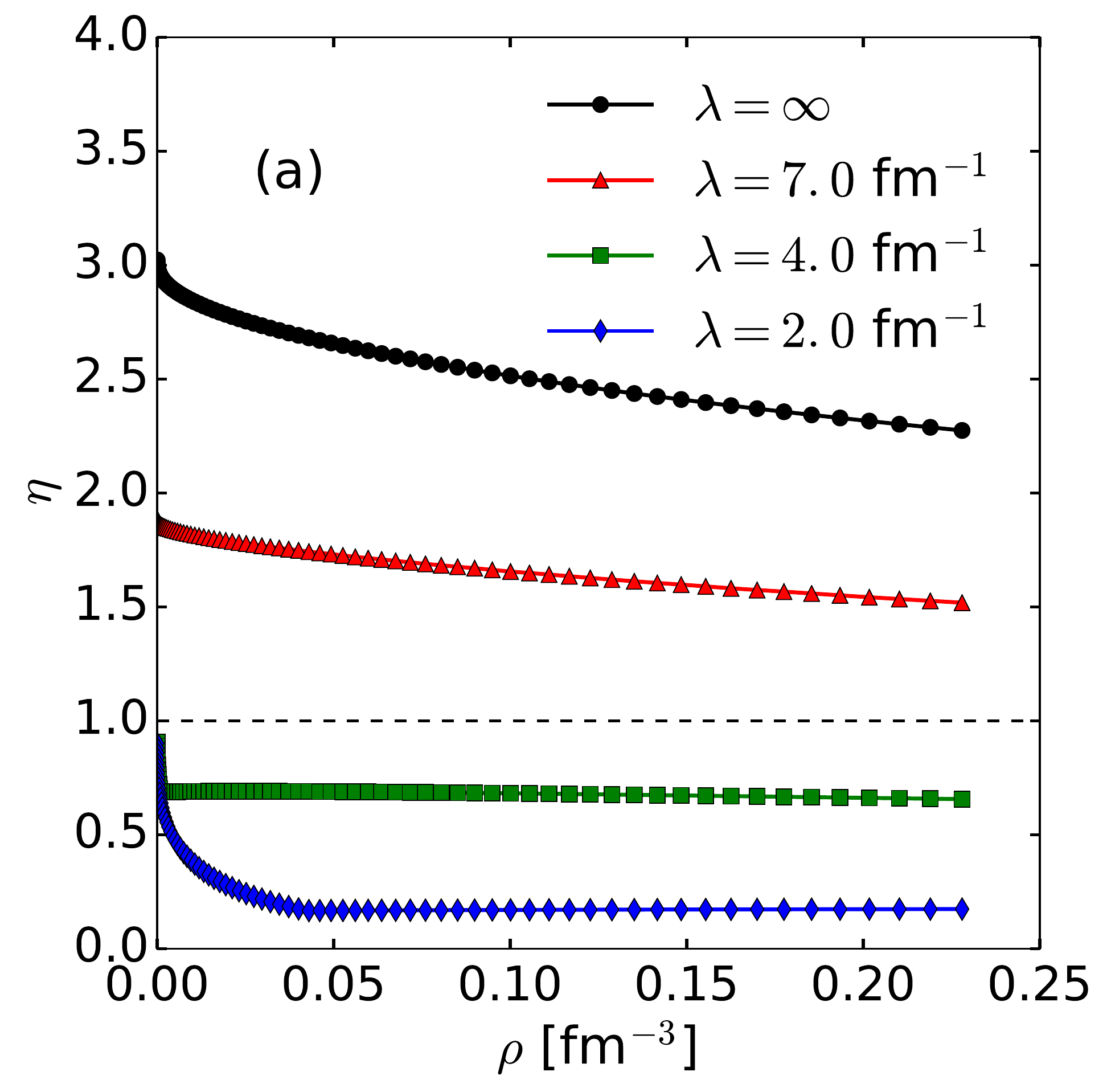}~
  \includegraphics[width=0.49\textwidth]{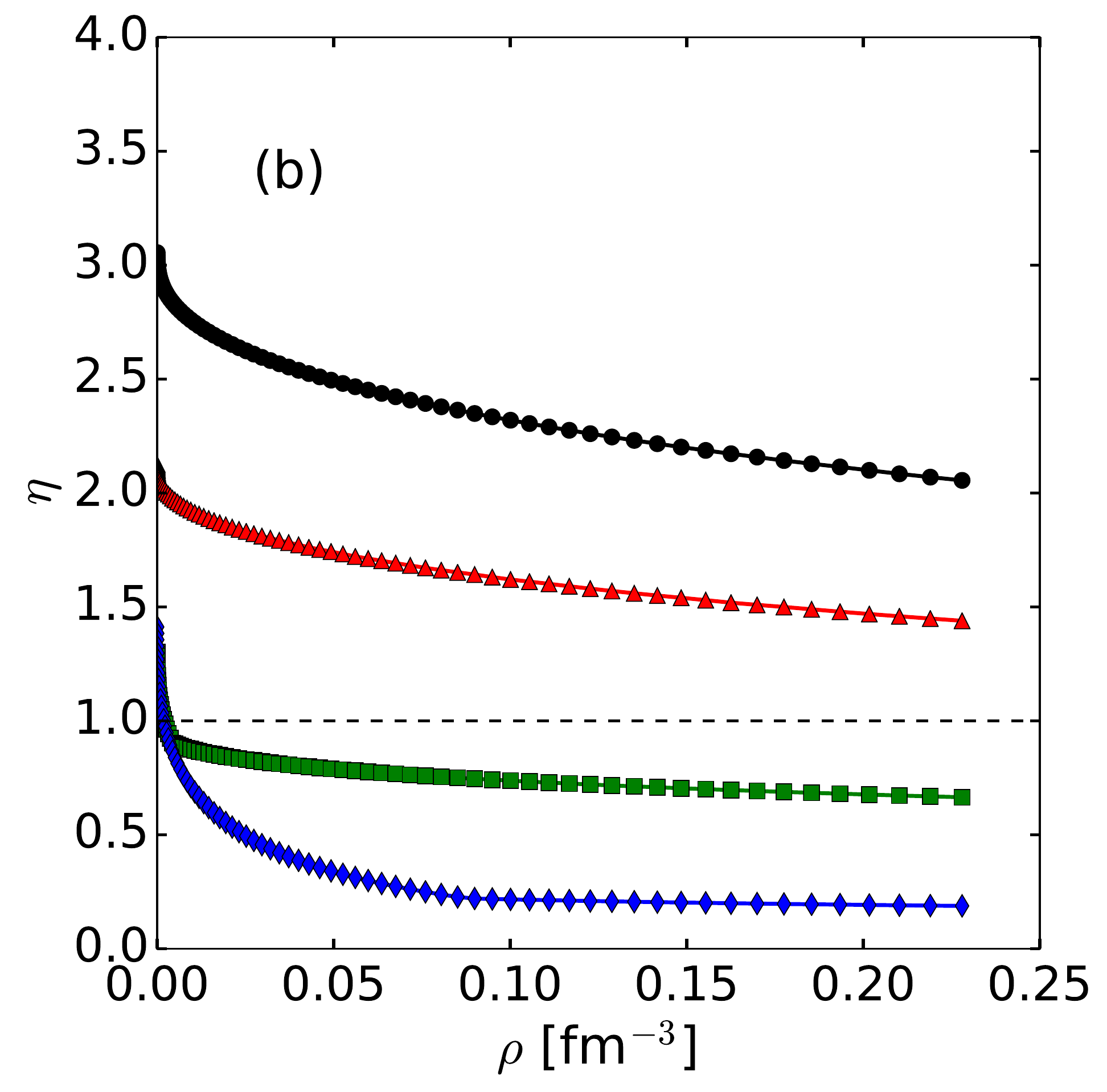}
  \caption{\textbf{(a)} The expansion parameter $\eta$ in Eq.~\eqref{eq:eta} is plotted as a function of density $\rho$ for the $\osz$ partial wave using the AV18 potential. 
  Four different SRG $\lambda$ scales are shown.
  \textbf{(b)} The same as (a) but for the $\tso$ partial wave.
  }
  \label{fig:pp_eign_AV18}
\end{figure}

\begin{figure}[t]
 \begin{center}
  \includegraphics*[width=2.4in]{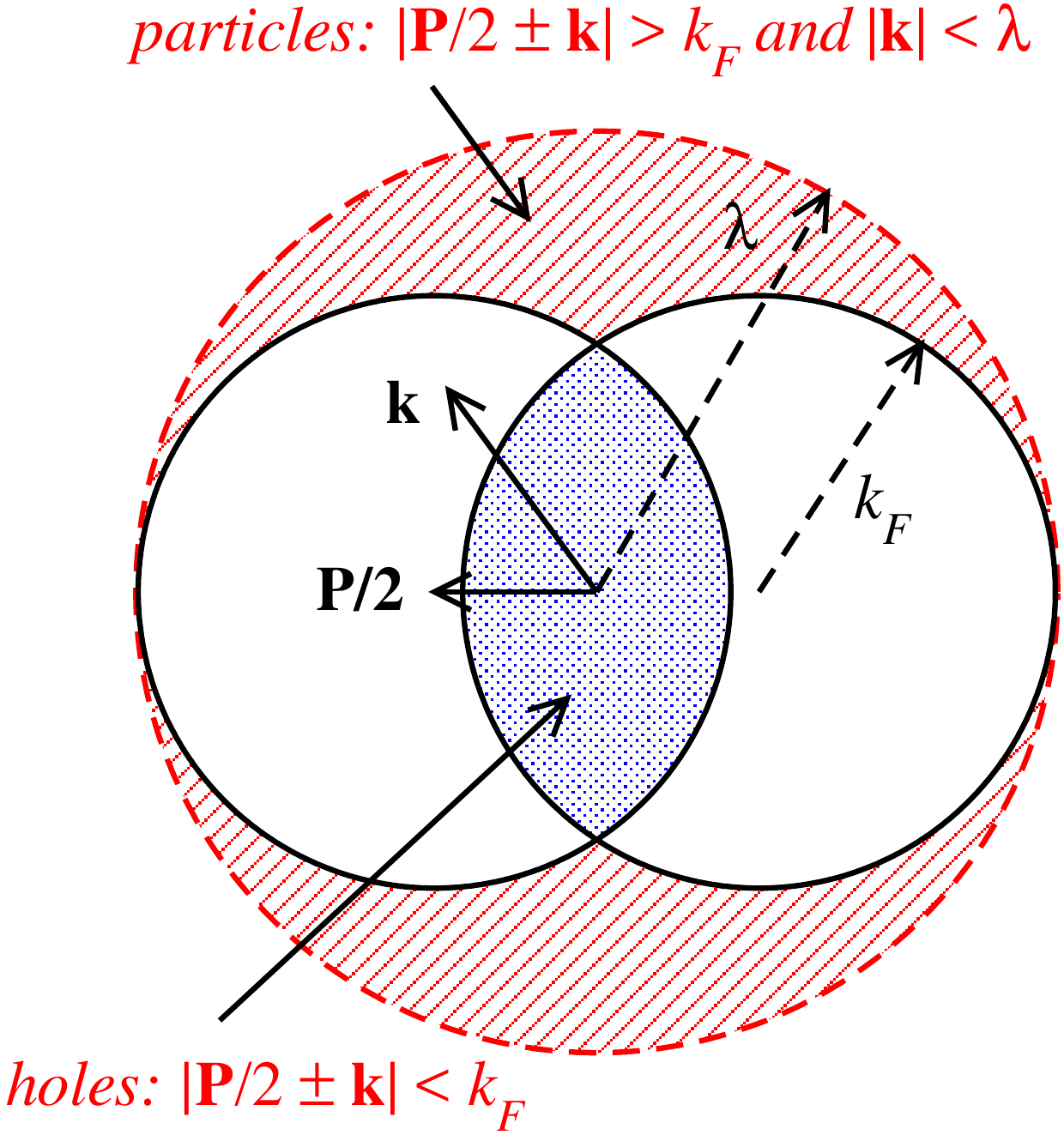}
 \end{center}
 \caption{Diagram of two Fermi spheres illustrating the hole and particle phase space  available for two-body interactions.
 $\lambda$ here serves as a scale for which the phase space above is effectively decoupled from the phase space below.
 This is in contrast to other methods (e.g., $\vlowk$) where the $\Lambda$ is an actual cutoff in the model space.}
 \label{fig:fermispheres}
\end{figure}

	In addition, Eq.~\eqref{eq:kernel_diag} allows for a clean and rigorous definition of potential perturbativeness; adding a rung to the ladder introduces an extra power of the kernel eigenvalue matrix $D$ and numerical prefactors.
	The dimensionless\footnote{$\eta$ can be seen to be dimensionless by noting that an extra rung also introduces an extra weight into Eq.~\eqref{eq:pp_est} for the interior particle momentum.} expansion parameter $\eta$ for the pp ladder is then simply,
\beq
	\eta \equiv \frac{2 m}{\pi \hbar^2} \;
	|\epsilon_{\text{max}}| \; , 
	\label{eq:eta}
\eeq
where $\epsilon_{\text{max}}$ is the largest eigenvalue of the kernel $\kernel$ and we take the absolute magnitude. 
	The potential is perturbative in the pp ladder if $\eta < 1$ and nonperturbative otherwise. 
	In Fig.~\ref{fig:pp_eign_AV18}, we plot $\eta$ against density for different values of the SRG scale $\lambda$ in the $\osz$ and $\tso$ partial waves. 
	Note that in the low density limit, irrespective of the SRG scale, the potential is nonperturbative in both waves, reflecting the fine-tuning in the two channels.
	For the unevolved potential $\lambda = \infty$, the potential is nonperturbative near saturation density $\rho \approx 0.16 \fm^{-3}$. 
	However as the flow parameter lowers, the potential below the scale $\lambda$ becomes effectively decoupled from the potential above. 
	As a result, the particle phase space becomes increasingly constrained and $\eta$ decreases, see Fig.~\ref{fig:fermispheres}. 
	By the time $\lambda = 4.0 \fmi$, $\eta$ is less than 1 and the pp channel is perturbative at saturation density for these waves.
	
	Our analysis here is closely related to the use of Weinberg eigenvalues that arise 
	in studying the convergence of the scattering Born series \cite{Hoppe:2017aa}.
	The Born expansion can be rendered as a geometric series with convergence being dictated by the eigenvalues of the operators,
\beq
	G_0 V | \xi \ra = \xi | \xi \ra \; ,
	\label{eq:weinberg_eigen}
\eeq
where $G_0$ is the non-interacting propagator.
	If the eigenvalues of the system are of order $1$ or greater, then the Born series does not converge. 
	The formulation of the pp kernel in Eq.~\eqref{eq:pp_kernel} looks similar to the above though with a more symmetric form,
\beq
	\sqrt{G_0} \ V \sqrt{G_0} 
	| \widetilde{\xi} \ra = 
	\widetilde{\xi} 
	| \widetilde{\xi} \ra\; .
\eeq
	Multiplying the left hand side of the above by $\sqrt{G_0}$ and defining 
	$ \sqrt{G_0} |\widetilde{\xi} \ra \equiv |\xi \ra$ brings it into the form of Eq.~\eqref{eq:weinberg_eigen}.
	Setting $\kav = 0$ in Eq.~\eqref{eq:pp_kernel} and working with a $\vlowk$ potential in the $\osz$ partial wave, our expansion parameter exactly tracks the largest Weinberg eigenvalue in Ref.~\cite{Ramanan:2007bb} (see Fig.~1). 
    The finite density results for $\eta$ given here use $\kav \neq 0$ and so will differ in general from 
    Weinberg eigenvalues in~\cite{Ramanan:2007bb}.

\section{Unitary Fermi Gas}
\label{sec:unitary}

An interesting nonperturbative extreme of a Fermi system is the unitary limit, where 
the scattering length $a$ of the system is taken to infinity and the inter-particle
 separation $\kf^{-1}$ is taken to be much larger than the effective range of the
  potential $r$~\cite{Georgini:2008}, 
\beq
	\text{Unitary Limit:}
	\qquad
	\lim_{a\to \infty} 
	\quad
	\text{and}
	\quad
	\kf r \ll 1	
	\;.
\eeq
	For $NN$ scattering, this limit serves as an approximate description for the $\osz$ and $\tso$ partial waves, as both of these channels have unnaturally large scattering lengths.
	Note that the low density values of $\eta$ in Fig.~\ref{fig:pp_eign_AV18} are very close to or 
	above 1, reflecting the lack of perturbative convergence.
	In this section we demonstrate that our averaging and factorization procedure explicitly reproduces the nonperturbativeness of the unitary gas.
	For simplicity we assume our potential $V$ is a pure contact with no momentum or spin dependence,
\beq
	V = C_0 \; .
\eeq
	For such a simple potential, the T matrix scattering amplitude is a summable geometric series given by~\cite{Kaplan:1998we},
\beq
	T(E) = \frac{C_0}{1 - I(E)} \; ,
\eeq
where $I(E)$ is a generic loop integral in the bubble chain,
\beq
	I(E) \equiv \int \frac{d^3\kvec}{(2\pi)^3}
	\frac{C_0}{E - \hbar^2 k^2 / m}	 \; .
\eeq
	In the unitary limit, the T matrix has a pole at zero energy, $I(0) = 1$. 
	Imposing a sharp momentum cutoff $\Lambda$ on this generic loop integral, the value for $C_0$ can be found,
\beq
	1 = \frac{1}{2\pi^2} \; C_0
	\int_0^{\Lambda} dk \; k^2 
	\frac{1}{0 - \hbar^2 k^2 / m}
	\quad
	\implies
	\quad
	C_0 = - \frac{2 \pi^2 \hbar^2}{\Lambda m} \; .
\eeq

	Working with relative and center-of-mass momentum variables in a single-particle basis, the energy per particle for the $n$th rung in the pp ladder using our two approximations in Eq.~\eqref{eq:nn_approx} is written as,
\begin{align*}
	\frac{E^{(n)}_{\rm pp}}{N} 
	\propto
	\left(
	\frac{1}{2}
	\right)^n
	\left(
	\frac{m}{\hbar^2}
	\right)^{n-1}
	(-1)^{n-1}
	\left(
	\frac{4\pi}{8 \pi^3}
	\right)^{n}
	\prod_{i}^n
	\tr_{\sigma_i \tau_i}
	\left[
	\left(
	1 - P_{12}
	\right)^n
	\right]
	\int 
	dk_i  \;
	K^{n-2}
	\numberthis
\end{align*}
where we have only included factors that scale with additional rungs in the ladder. 
	The factor $1/2$ comes from the symmetry of equivalent lines, $m/\hbar^2$ from energy denominators, $(-1)$ from flipping the terms in the energy denominators, $4\pi/8 \pi^3$ from angular integrations, and $\kernel$ is again the pp kernel in Eq.~\eqref{eq:pp_kernel} in the continuum limit. 
	The $n$th term of the spin-isospin trace factors can be written by noting that $(1 - P_{12})^2 = 2 (1 - P_{12})$ such that,
\beq
	\tr_{\sigma_i \tau_i} 
	\left[
	\left(
	1 - P_{12}
	\right)^n
	\right]
	=
	12 \times 2^{n-1} \; ,
\eeq
	with $2^{n-1}$ canceling the $\left(1/2\right)^n$ scaling from the symmetry factors.
	Furthermore as the unitary limit also implies that $\lambda \propto r^{-1}$ meaning that $\lambda \gg \kf$, the energy denominators and particle Pauli blockers can be expanded in a series where to leading order,
\beq
	\Qavp \to 1 \; \quad
	\text{and}
	\quad
	\frac{1}{k_a^2 - \kav^2} \to 
	\frac{1}{k_a^2} \; ,
	\quad
	\text{resulting in}
	\quad
	\sqrt{\frac{\Qavp (k_a, \Pav) \; k_a^2}
	{k_a^2 - \kav^2}}	 \to 1 \; , 
\eeq
such that the kernel $\kernel$ is equivalent to the potential, $\kernel = V$. 
	Therefore, counting factors that contribute with adding an additional rung to the ladder results in the expansion parameter being,
\begin{align*}
	\eta \; = \; \frac{m}{\hbar^2}
	\;
	(-1)
	\;
	\frac{4\pi}{8\pi^3}
	\;
	C_0
	\int_0^\Lambda dk \; = \; 1 
	\numberthis
\end{align*}
and the system is nonperturbative as expected.
	The above line of argument can also be used when the potential is treated as separable~\cite{Kohler:2007}, a good approximation for low-momentum potentials.

	This analysis is also consistent with the Weinberg eigenvalues of the system in free space.
	For positive Weinberg eigenvalues associated with bound or near bound states, the values are of order 1 indicating the non-convergence of the Born series.
	As this is relevant physics that does not depend on resolution, these eigenvalues do not flow with the RG scale in free space, see Fig.~3 in Ref.~\cite{Bogner:2005sn}.
	Setting $\kav = 0$ and $\Qavp = 1$ in Eq.~\eqref{eq:pp_kernel} for the $\osz$ and $\tso$ partial waves yields $\eta \sim 1$ or greater for the different SRG scales as expected from the large $NN$ scattering lengths.
	Likewise setting $\kav^2 = B_d m / \hbar^2$ in Eq.~\eqref{eq:pp_kernel}, where $B_d$ is the deuteron binding energy, explicitly reproduces the deuteron pole ($\eta = 1$) when the repulsive Weinberg eigenvalue is less than 1, see Fig.~4 in Ref.~\cite{Bogner:2005sn}.

\section{Three-Body Forces}
 \label{sec:three_body}

	In this section, we discuss estimates for a $3N$ contact in uniform matter and the complementary normal-ordered $NN$ force.
 
\subsection{Three-Body Contact and Normal Ordering}
 \label{sec:no3b}

	The preceding discussion only estimated contributions from two-body interactions.
	However Hamiltonians from $\eft$ will have three-body and higher operators, with three-body forces first appearing at \NNLO~in the $\Delta$-less chiral expansion~\cite{VanKolck:1994yi,
	Epelbaum:2002vt}.
	Matrix elements for a three-body operator $\vnnn$ in the single-particle basis are,
\beq	
	\la 1' 2' 3' | \vnnn \anti | 1 2 3 \ra \; ,
\eeq
where $\anti$ is the antisymmetrizer and we use the short hand $|1 \ra = | \pvec_1 \s_1 \is_1 \ra$.
	For simplicity, here we only consider the pure $3N$ contact term $\ve$, 
\beq
	\ve = \frac{c_E}{2 \fpi^4 \lchi}
  \sum_{i \neq j} \is_i \cdot \is_j \; ,
  \label{eq:3N_con}
\eeq
and set the constant to unity, $c_E = 1$.

	Like two-body forces, three-body forces must also be regulated when solving for three-body LECs via Faddeev equations.
	A common choice is a nonlocal regulator of the form~\cite{Epelbaum:2002vt},
\begin{align*}
	f(\pvec_1, \pvec_2, \pvec_3) \equiv 
	\exp \left[ - 
	\left(\frac{ p_1^2 + p_2^2 + p_3^2
	- \pvec_1 \cdot \pvec_2 - \pvec_2 \cdot \pvec_3 - \pvec_1 \cdot \pvec_3}
	{3 \threecut^2} \right)^n 
	\right] \; , 
	\numberthis
  \label{eq:nonlocal_reg}
\end{align*}
with $n$ some integer (we choose $n=2$ hereafter) and $\threecut$ the $3N$ cutoff. 
	This regulator has the particular advantage in that it is invariant under permutation symmetry, which can be easily seen by applying $P_{ij}$ to Eq.~\eqref{eq:nonlocal_reg} for any $i$ and $j$. 
	Both the incoming and outgoing momenta are regulated such that the potential $\vnnn$ is,
\beq
	\vnnn  
	\xrightarrow{\text{reg.}}
	f(\pvec_1', \pvec_2', \pvec_3') \;
	\vnnn  \;
	f(\pvec_1, \pvec_2, \pvec_3) \; .
\eeq
	Eq.~\eqref{eq:nonlocal_reg} can also be rewritten in two different ways~\cite{Hebeler:2009iv}:
\bseq
\beq
	f(k,j) =
	\exp \left[ - 
	\left(\frac{k^2 + 3j^2/4}
	{\threecut^2}
	\right)^2
	\right] \; ,
\label{eq:nonlocal_reg_jac}
\eeq
\beq
	f(\Pvec, \pvec_3, k) = 
	\exp \left[ - 
	\left(\frac{P^2/4 + 3 k^2 + p_3^2 - \Pvec \cdot \pvec_3}
	{3 \threecut^2}
	\right)^2
	\right] \; , 
\label{eq:nonlocal_reg_dd}
\eeq
\eseq
where 
$\kvec = \frac{1}{2} (\pvec_1 - \pvec_2)$ and 
$\jvec = \frac{1}{3} (2 \pvec_3 - \pvec_1 - \pvec_2)$ are Jacobi momenta and $\Pvec = \pvec_1 + \pvec_2$ is the center-of-mass momentum in the 1,2 subsystem.

	It is common to reorganize the vacuum three-body forces where, in the language of second quantization, the three-body creation and annihilation operators are normal-ordered with respect to a reference state 
\cite{PhysRevC.76.034302,PhysRevC.81.024002,Bogner:2009bt}.
	A common approximation in many ab-initio approaches~\cite{PhysRevLett.109.052501,
	PhysRevC.90.041302,
	Soma:2012zd,
	PhysRevLett.113.142502} is the so-called normal-ordered two-body (NO2B) approximation where after normal-ordering, only two- and lower-body forces are kept for reasons of computational efficiency. 
	The resulting two-body term is given by, for our state $\refstr$,
\begin{align*}
	\la 1' 2' | \vnnnd |1 2 \ra
	&=
	\tr_{\s_3}
	\tr_{\is_3}
	\int 
	\frac{d^3 \pvec_3'}{(2\pi)^3}  
	\frac{d^3 \pvec_3}{(2\pi)^3}
	(2\pi)^3 \delta^3(\pvec_3 - \pvec_3')
	\; n(\pvec_3)
	\\ 
	&\times 
	\la 1' 2' 3' | \vnnn 
	\left(
	1 - P_{13} - P_{23}
	\right)	
	| 1 2 3 \ra \; .
	\numberthis
	\label{eq:gen_normal}
\end{align*}
Note that Eq.~\eqref{eq:gen_normal} is not yet antisymmetrized with respect to particles 1 and 2. 
	Applying Eq.~\eqref{eq:gen_normal} to the $3N$ contact in Eq.~\eqref{eq:3N_con} yields,
\beq
	\ved = - \frac{6 \; c_E}
	{\fpi^4 \lchi} 
	\rho_f (\Pvec, k, k') \; ,
	\label{eq:dd_contact}
\eeq
where $\rho_f$ is the integration over the averaged single-particle momentum with the uncorrelated Fermi-Dirac distribution function and the regulators,
\beq
	\rho_f (\Pvec, k, k') = \int 
	\frac{d^3\pvec_3}{(2\pi)^3}
	n(\pvec_3) 
	f(\Pvec, \pvec_3, k) 
	f(\Pvec, \pvec_3, k') \;,
	\label{eq:rhof}
\eeq
	and we have used the fact that the total momentum is conserved.
	Note that in averaging over the presence of the third particle, we implicitly defined a preferred frame, namely the rest frame of the non-interacting Fermi sea. 
	This results in the effective force $\ved$ gaining explicit dependence on the center-of-mass momentum $\Pvec$ for the two scattering particles.

\subsection{Normal Ordered Terms at Second Order}

\begin{figure}[tbh]
 \begin{center}
  \includegraphics[width=0.3\textwidth]{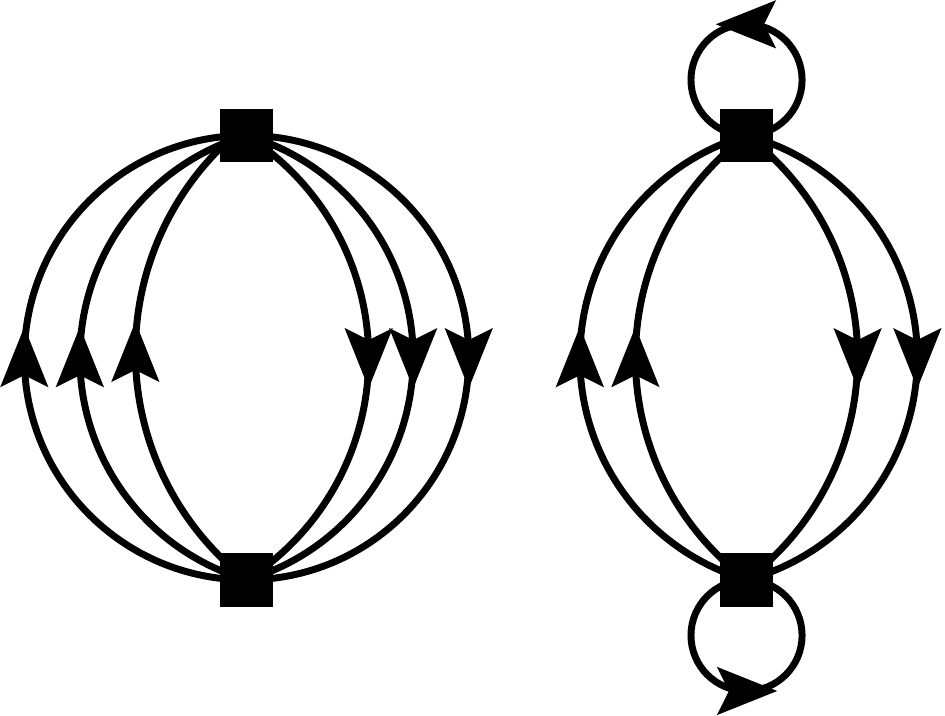}
 \end{center}
 \caption{Goldstone diagrams of the second order energy for $\ve$ and $\ved$.}
 \label{fig:2nd_order_chiral}
\end{figure}

	In this section we compute the second-order energy contributions of the effective two-body potential from normal ordering, also called the density-dependent (DD) term, and the remaining three-body piece, also called the residual (RE) term.
	In general, the energy contributions from these diagrams will be scale and scheme dependent and therefore the validity of the NO2B approximation depends on a choice of regularization and cutoff.

	In the single-particle basis, the second-order energy per particle of the DD two-body term and RE three-body term are given by, respectively,

\bseq
\begin{align*}
	\frac{\edd}{N} = 
	&\frac{1}{4 \rho}
	\left(
	\prod_{i=1}^4 
	\tr_{\s_i}
	\tr_{\is_i}
	\int \frac{d^3 \pvec_i}{(2\pi)^3}
	\right)
	\frac{\la 1 2 | \ved P_{12} | 3 4 \ra 
	\la 34 | \ved P_{12} | 1 2 \ra}
	{p_1^2 + p_2^2 - p_3^2 - p_4^2} 
	\\
	&\times
	n(\pvec_1) n(\pvec_2)
	\nb (\pvec_3) \nb (\pvec_4) \;
	(2\pi)^3 \delta^3(\pvec_1 + \pvec_2 - \pvec_3 - \pvec_4)
	\numberthis
	\;,
\end{align*}
\begin{align*}
	\frac{\eres}{N} = 
	&\frac{1}{36 \rho}
	\left(
	\prod_{i=1}^6
	\tr_{\s_i}
	\tr_{\is_i}
	\int \frac{d^3 \pvec_i}{(2\pi)^3}
	\right)
	\frac{\la 1 2 3| \ve \anti | 4 5 6 \ra 
	\la 4 5 6 | \ve \anti | 1 2 3\ra}
	{p_1^2 + p_2^2 + p_3^2 - p_4^2 - p_5^2 - p_6^2} \;
	\\
	&\times
	n(\pvec_1) n(\pvec_2) n(\pvec_3)
	\nb (\pvec_4) \nb (\pvec_5) \nb (\pvec_6) \;
	(2\pi)^3 \delta^3(\pvec_1 + \pvec_2 + \pvec_3 - \pvec_4 - \pvec_5 - \pvec_6)
	\numberthis
	\;.
\end{align*}
\eseq
	Goldstone diagrams for these two expressions are given in Fig.~\ref{fig:2nd_order_chiral}.
	Converting to relative and center-of-mass coordinates, noting that the spin-isospin traces factorize to give numerical prefactors\footnote{The factors are 24 for the density-dependent diagram and 144 for the residual diagram.}, and simplifying we find, 
\begin{align*}
	\frac{\edd}{N} &= 
	\left(\frac{6 \; c_e}{\fpi^4 \lchi} \right)^2
	\frac{6}{\rho}
	\int \frac{dk \; dk' \; d^3\Pvec}{32 \pi^7}
	k^2 \; k'^2 \;
	\frac{\rho_f^2 (\Pvec, k, k')}{k^2 - k'^2} \;
	\Qavp (P, k'; \kf) \;
	\Qavh (P, k ; \kf) \; ,
	\numberthis
	\label{eq:energy_no2b}
\end{align*}
for the density-dependent term, where the angular integrals over $\khat$ and $\khat'$ have been done exactly, and 
\begin{align*}
	\frac{\eres}{N} =
	&\left(\frac{c_e}{2 \fpi^4 \lchi} \right)^2 
	\frac{4}{\rho}
	\int \frac{d^3 \kvec \; d^3 \kvec' \; 
	d^3 \jvec \; d^3 \jvec'
	d^3 \Wvec}
	{(2\pi)^{15}} 
	\frac{f(k,j)^2 f(k',j')^2 \;}
	{
	k^2 + \frac{3}{4} j^2 - k'^2 - \frac{3}{4} j'^2
	}
	\\
	\times \;
	&n(\Wvec/3 + \jvec) \;
	\nb(\Wvec/3 + \jvec')  \;
	\Qh (\Wvec/3 - \jvec/2, \kvec; \kf) \;
	\Qp (\Wvec/3 - \jvec'/2, \kvec'; \kf) \; ,
	\label{eq:energy_res}
	\numberthis
\end{align*}
for the residual where $\Wvec = \pvec_1 + \pvec_2 + \pvec_3$ is the center-of-mass momentum of 
the 3 particle system.
	As in the $NN$ sector, the Pauli operators involving the Jacobi momenta can be angle-averaged giving,
\begin{align*}
	\frac{\eres}{N} =
	\left(\frac{c_e}{2 \fpi^4 \lchi} \right)^2
	\frac{4}{\rho}	
	&\int \frac{dk \; dk' \; 
	dj \; dj' \;
	dW}
	{32 \pi^{10}}
	k^2 \; k'^2 \; j^2 \; j'^2 \; W^2
	\frac{f(k,j)^2 f(k',j')^2 \;}
	{
	k^2 + \frac{3}{4} j^2 - k'^2 - \frac{3}{4} j'^2
	}
	\\
	&\times
	\Qtp (W, j', k') \; \Qth (W, j, k)
	\numberthis
	\;,
\end{align*}	
where the derivation and functional forms of the three-body hole $\Qth (W, j, k)$ and particle $\Qtp (W, j', k')$ angle-averaged operators are given in Appendix~\ref{sec:three_body_fermi_spheres}.
	For three-body potentials without angular dependence like our three-body contact, this procedure is exact. 
	The accuracy of applying the angle-average approximation to three-body potentials with pion exchange is an open question. 
 
	In making estimates for the DD diagram we assume, as in the $NN$ case, that the particle relative momentum is sufficiently larger than the total and hole relative momentum, $k' \gg P, k$. 
	This leads to the same approximations employed in Eq.~\eqref{eq:nn_approx}.
	Also we set the total momentum to zero ($P=0$) in the function $\rho_f$ as $P \sim \kf$ and its effect in the exponential of $\rho_f$ will be small. 
	This approximation has been investigated previously and shown to be quite accurate, see e.g., Refs.~\cite{Hebeler:2009iv,
	PhysRevC.81.024002,
	PhysRevC.94.054307}.
	These approximations then yield,
\begin{align*}
	\frac{\edd}{N} &\approx 
	\left(\frac{6 \; c_e}{\fpi^4 \lchi} \right)^2
	\frac{6}{\rho}
	\int \frac{dk \; dk'}{8 \pi^6}
	k^2 \; k'^2 \;
	\frac{\rho_f^2 (0, k, k')}{\kav^2 - k'^2} \;
	\Qavp (\Pav, k'; \kf) \;
	\hhps \left( \frac{k}{\kf}, \kf \right) \; .
	\numberthis
	\label{eq:dd_approx}
\end{align*}

\begin{figure}[t]

  \includegraphics[width=0.54\textwidth]{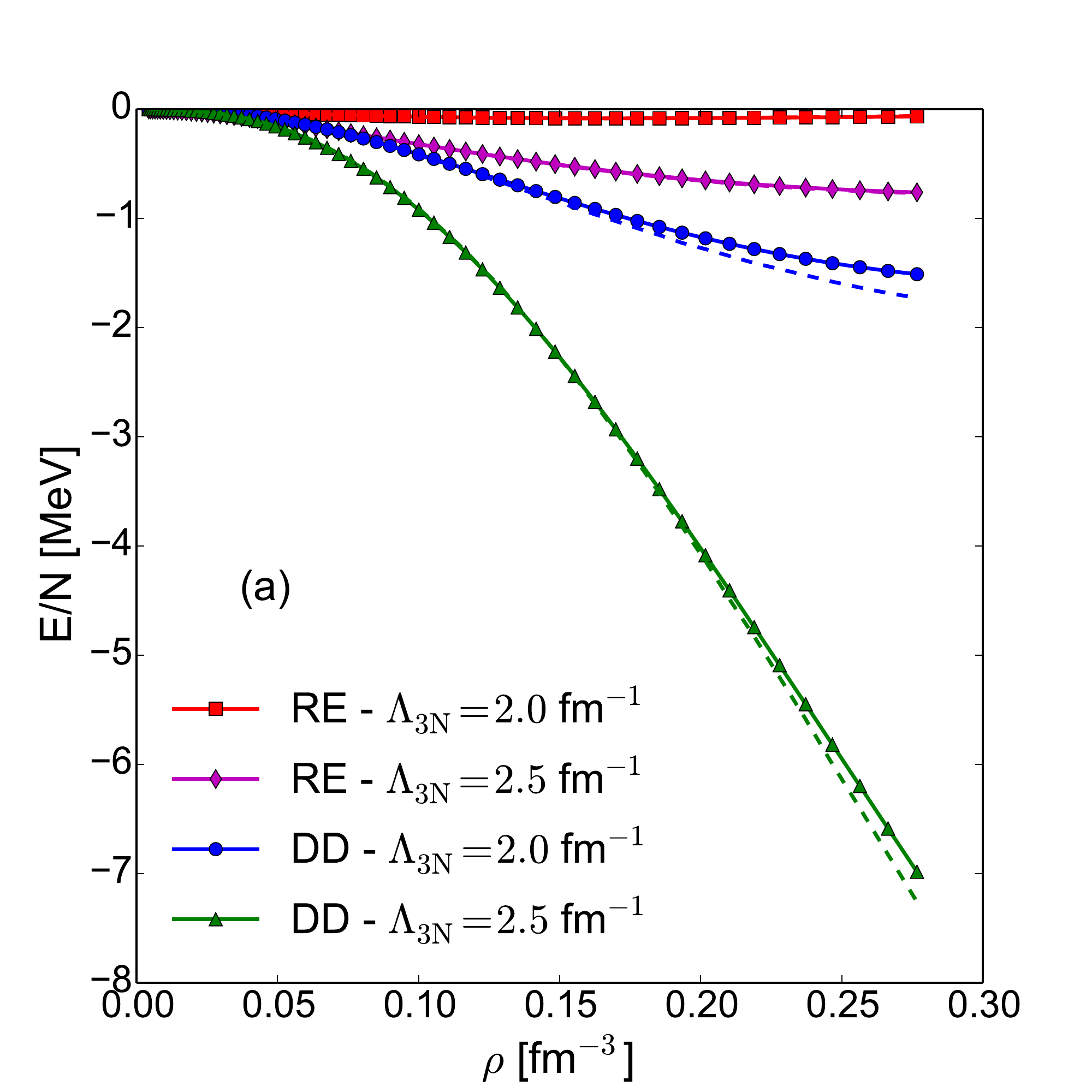}~
  \hspace*{-0.3in}
  \includegraphics[width=0.54\textwidth]{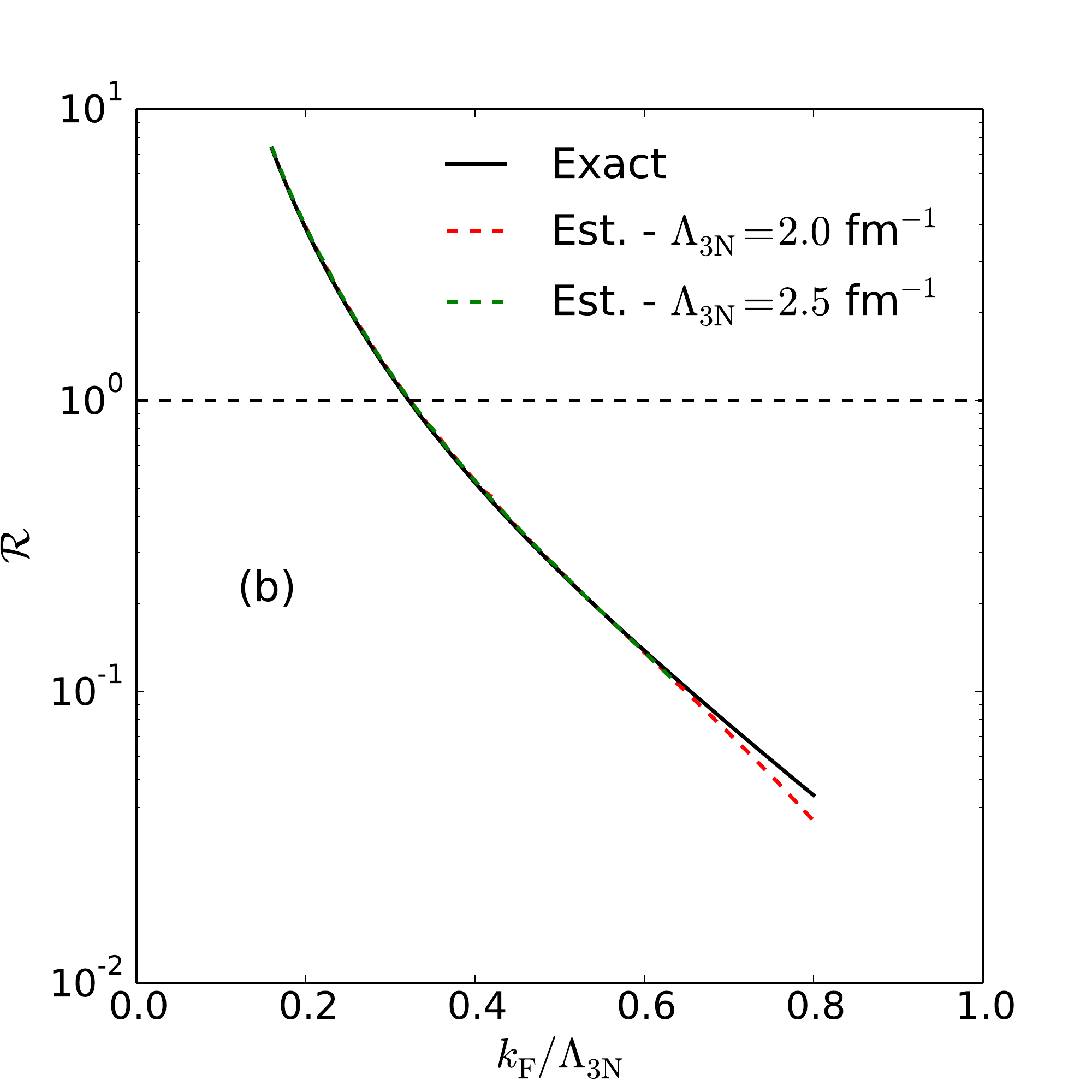}

  \caption{
  \textbf{(a)} The second-order energy per particle in nuclear matter is plotted as a function of density for the density-dependent (DD) and residual (RE) diagrams of the $3N$ contact $V_E$ at two different cutoffs $\threecut$. 
  Both exact (solid) and estimates (dashed) are shown for each diagram and cutoff. 
 \textbf{(b)} The dimensionless ratio in Eq.~\eqref{eq:noratio} is plotted as a function of the Fermi momentum over the $3N$ cutoff. 
  }
  \label{fig:three_body_avg}
\end{figure}

	For the RE diagram, we assume the integrand is dominated by regions where the particle Jacobi momenta is sufficiently larger than the total and the hole Jacobi momenta, $k', j' \gg W, j, k$. 
	Analogously to the $NN$ sector, the hole Jacobi momenta in the energy denominator and the total momentum $W$ in the particle Pauli blocker are replaced with their RMS averages,
\begin{align*}
	\frac{\eres}{N} \approx
	\left(\frac{c_e}{2 \fpi^4 \lchi} \right)^2
	\frac{4}{\rho}	
	&\int \frac{dk \; dk' \; 
	dj \; dj' \;
	dW}
	{32 \pi^{10}}
	k^2 \; k'^2 \; j^2 \; j'^2 \; W^2
	\frac{f(k,j)^2 f(k',j')^2 \;}
	{
	\kav^2 + \frac{3}{4} \jav^2 - k'^2 - \frac{3}{4} j'^2
	}
	\\
	&\times
	\Qtp (\Wav, j', k') \; \Qth (W, j, k)
	\label{eq:res_approx}
	\numberthis
	\;,
\end{align*}	
where the RMS values for $j$ and $W$ are, 
\beq
	\jav = \sqrt{\frac{2}{5}} \kf \; ,
	\qquad
	\Wav = \sqrt{\frac{9}{5}} \kf \; .
\eeq
	As a result of the approximations applied in Eq.~\eqref{eq:res_approx}, the integrations over the hole and particle phase space factorize.
	The hole and particle phase space in Eq.~\ref{eq:dd_approx} also nearly factorize with slight residual coupling in the $\rho_f$ function of Eq.~\eqref{eq:rhof} via integrating the two exponentials in Eq.~\eqref{eq:nonlocal_reg_dd} over the single-particle hole state.
	Figure~\ref{fig:three_body_avg} shows plots of the exact and estimated values of the second-order energy per particle at two commonly used $3N$ cutoffs.
	For both cutoffs, the estimates do an excellent job of reproducing the exact values with absolute errors in the DD diagram of $0.05$ and $0.02$ MeV for $\threecut = 2.0$ and $2.5 \fmi$ respectively near saturation. 
	These results confirm that our approximations are well motivated and capturing the relevant physics.
	
	A figure of merit for the validity of the NO2B approximation is the dimensionless ratio, 
\beq	
	\noratio 
	\equiv \frac{\eres}{\edd} \; ,
	\label{eq:noratio}
\eeq
where the ratio terms can either be exact values or our estimates.
	Figure~\ref{fig:three_body_avg} shows this ratio as a function of the dimensionless quantity $\kf / \threecut$ both in the exact case and for our estimates at two different cutoffs. 
	The exact ratio is very well reproduced by our estimates with $\noratio \sim 1$ at around a value of $\kf / \threecut \sim 0.33$. 
	The behavior of the ratio $\noratio$ can be understood by noting that at low $\kf / \threecut$, a good deal of phase space exists between the Fermi surface and the cutoff. 
	As the RE diagram has 3 particles and 3 holes compared to the DD's 2 particles and 4 holes, the extra particle phase space increases the comparative importance of the residual term.
	As $\kf / \threecut$ is raised, the particle phase space becomes constrained\footnote{In an analogous way to Fig.~\ref{fig:fermispheres} but with three Fermi spheres.} decreasing the residual energy.
	For comparison, dimensional regularization with minimal subtraction as done in Ref.~\cite{Kaiser:2012} gives $\noratio \approx 1/2$.
	Note that Fig.~\ref{fig:three_body_avg} only shows this ratio for the regulator choice in Eq.~\eqref{eq:nonlocal_reg}; local regulators with similar cutoffs in nuclear matter give much larger residual terms~\cite{PhysRevC.89.014319} cf., Ref.~\cite{Dyhdalo:2016ygz}.
	Although $3N$ potentials with momentum dependence will modify the ratio plot in Fig.~\ref{fig:three_body_avg}, it serves as a quantitative starting point for the relative importance of the DD and RE terms. 
	Furthermore, the accuracy of our estimates support a phase space approach for the $3N$ diagrams as well as extensions to higher orders in MBPT.

\section{Conclusion}
\label{sec:conclusion}

	Chiral potentials with soft cutoffs and renormalization group approaches have resulted in nuclear potentials much more amenable to perturbative approaches,  but a systematic power counting is lacking. 
	In this work, we revisited power counting in nuclear matter with the AV18 potential softened using the SRG. 
	Utilizing physically motivated approximations based on phase space, we factorized momentum integrals appearing in individual terms of the pp and hh ladders.
	These approximations yield high fidelity estimates of the energy per particle in nuclear matter for different diagrams as well as expansion parameters for each channel.
	We then briefly showed why our analysis does not imply perturbativeness in the unitary limit.
	For $3N$ forces, a pure contact at second-order in MBPT was also considered along with the validity of the NO2B approximation. 
	
	Our $NN$ estimates were applied solely for the AV18 potential in the $\osz$ and $\tso$ partial waves.
	For the two SRG scales $\lambda = 4.0$ and $2.0 \fmi$, the absolute difference between an exact calculation and our estimates near saturation is, at worst, a few tenths of an MeV per particle going up to fourth-order.
	These channels were also found to be perturbative in the pp ladder starting around the SRG scale $\lambda=4.0 \fmi$. 
	For both the unevolved potential and all considered SRG scales, the hh channel was found to be perturbative. 
	Starting around $\lambda = 2.0 \fmi$, terms in the hh ladder were found to be comparable in importance to terms in the pp ladder near saturation density.
	This reinforces previous suggestions~\cite{Hebeler:2010xb,Bogner:2005sn} that performing perturbation theory in the \textit{potential} itself may be sufficient for softened interactions.
	Although we have confined our $NN$ discussion to AV18 to illustrate the efficacy of our estimate formalism and the onset of perturbativeness for a more traditional hardcore potential, the same analysis can be applied to potentials from $\eft$.
	Because of the flow to universal potentials that is well realized by $\lambda = 2.0 \fmi$,
	the numerical results from $\eft$ for this $\lambda$ and below will be the same
	as for AV18.
	
	In the $3N$ sector, we also examined the simplest interaction at second-order in MBPT, a spin-independent contact term. 	
	This interaction was normal ordered with respect to our finite density reference state to produce an effective two-body force. 
	Both the two-body and residual three-body force were calculated at second-order and compared to estimates.
	The estimates were found to closely reproduce the energy per particle in nuclear matter as well as the ratio of the two second-order terms.
	The NO2B approximation was then shown, for our simple interaction, to break down in the vicinity of $\kf / \threecut \sim 0.33$. 
	
	Our analysis does not directly extend to the particle-hole channel, as the results in 
	Appendix~\ref{sec:pw_appendix} do not apply and the partial waves do not factorize.
	Assessing the ultimate size of particle-hole contributions will be crucial to a systematic power counting for softened interactions.
	Other topics to be studied include explorations of power counting with novel 
	SRG generators~\cite{Li:2011sr}, one-body potentials, chiral effective field theory potentials, and around different
	reference states.
	We want to understand the
	impact of more complicated energy spectra and SRG evolution on $3N$ forces, as well as the scaling of higher many-body forces.
	Work on these fronts is ongoing. 

\begin{acknowledgments}
	We thank K. Hebeler and C. Drischler for interesting discussions and for providing numerical values for comparison. 
	This work was supported in part by the National Science Foundation under Grant Nos. PHY-1306250, PHY-1404159, and PHY-1614460, and the NUCLEI SciDAC Collaboration under DOE Grants DE-SC0008533 and DE-SC0008511

\end{acknowledgments}

\bibliography{refs_power_counting}% Produces the bibliography via BibTeX.

\clearpage
\newpage

\appendix

\section{Rules for Goldstone Diagrams}
 \label{sec:rules}

Here we list the rules for Goldstone diagrams with no one-body potentials, see e.g.,~\cite{baldo1999nuclear,Shavitt:2009}:

\begin{enumerate}
\item Integrate and/or sum over all internal momenta and spin-isospin degrees of freedom. 
\item Upward (downward) arrows designate particle (hole) states.
	The momentum magnitudes of these states satisfy the distribution functions,  
\beq
	n (\pvec_{\text{hole}})
	\; ,
	\quad
	\nb (\pvec_{\text{part.}})
	\equiv 
	1 - n (\pvec_{\text{part.}})
	\; ,
\eeq
for holes and particles respectively where $n(\pvec)$ is the usual Fermi-Dirac distribution at zero temperature,
\beq
	n (\pvec) \equiv
	\Theta (\kf - |\pvec|) \; .
\eeq
	\item Lines which close on themselves are counted as holes. 
	\item A potential interaction corresponds to a dashed line or vertex. Each is of the form:
\beq
	\la a b | \vnn (1 - P_{12}) | c d \ra \; ,
	\quad
	\la a b c | \vnnn \mathcal{A}_{123} | d e f \ra \; ,
	\quad 
	\ldots
	\; ,
\eeq
where the labels on the right enter and the lines on the left leave the interaction. Note that all interactions are antisymmetrized such that one diagram describes both direct and exchange terms. 
	\item Between successive vertices, there exist an energy denominator of the form,
\beq
\frac{Q}{E_0 - H_0} = \frac{Q}{\sum E_h - \sum E_p} \; ,
\eeq
where $Q$ is a Pauli blocking operator that enforces the requirements of rule 2.
	$E_p$, $E_h$ are the energies of particles and holes respectively. 
	Note that $E_p$, $E_h$ are not only free kinetic energies but also include self-energy terms.
\item An overall minus sign of the form,
\beq
(-1)^{h + l} \; ,
\eeq
where $h$ is the number of hole lines and $l$ is the number of closed loops.
\item Include a factor of $\frac{1}{n!}$ for each set of $n$ equivalent lines. Lines are equivalent if they begin and end at the same interaction and go in the same direction.
\end{enumerate}

\section{Potentials and ladders for \texorpdfstring{$NN$}{NN} interactions}
\label{sec:pw_appendix}

\subsection{Partial-Wave Basis}

	A given two-body energy contribution in MBPT has the generic form of,
\beq
	\sum_{\alpha \beta \ldots \chi \omega}
	\la \alpha | \vnna | \beta \ra \ldots 
	\la \chi | \vnna | \omega \ra \; ,
\eeq
where $\alpha$, $\beta$, $\ldots$, $\chi$, $\omega$ are our basis states with a complete set of quantum numbers and $\vnna$ is our anti-symmetrized $NN$ potential.
	As our goal is to evaluate energy contributions, all quantum numbers are summed (integrated) over.
	A two-body $NN$ state $| \alpha \ra$ expressed in a single-particle basis in momentum representation is given by a product state,
\beq
  | \alpha \ra = 
  | \pvec_1 \pvec_2 \ra \otimes 
  | \sigma_1 \sigma_{1z} ;
  \sigma_2 \sigma_{2z} \ra \otimes
  | \tau_1 \tau_{1z} ; \tau_2 \tau_{2z} \ra \; , 
\eeq
where $\pvec_i$ is the single-particle momentum of nucleon $i$, $\sigma_i$ and $\sigma_{iz}$ are the spin of nucleon $i$ and the spin projection along the quantization axis, and $\tau_i$ and $\tau_{iz}$ is the isospin of nucleon $i$ and the isospin projection along the quantization axis.
	However, in order to use our SRG evolved potentials, it is necessary to work instead with elements in a partial wave basis. 

	First, the single-particle momentums $\pvec_i$ are converted to relative momentum $\kvec$ and center-of-mass momentum $\Pvec$,
\beq
	\kvec = \frac{\pvec_1 - \pvec_2}{2}
	\; ,
	\qquad
	\Pvec = \pvec_1 + \pvec_2 \; .
\eeq
	The spins of the two nucleons can also be coupled to the total spin $S$ and total spin projection $S_z$ of the $NN$ state via,
\beq
	 | \s_1 \s_{1z} ;
 	 \s_2 \s_{2z} \ra
 	 =
 	 \sum_{S S_z}
 	 \C_{\s_1 \s_{1z} \s_2 \s_{2z}}^{S S_z}
 	 | S S_z \ra 
 	 \qquad 
 	 \text{where}
 	 \qquad
 	 \C_{\s_1 \s_{1z} \s_2 \s_{2z}}^{S S_z}
 	 =
 	 \la S S_z | \s_1 \s_{1z} ;
 	 \s_2 \s_{2z} \ra \; ,
\eeq
and the $\C$ terms are Clebsch-Gordan (CG) coefficients with the usual restrictions on the sums 
$| \s_1 - \s_2 | \leq S \leq | \s_1 + \s_2 |$ 
and $-S \leq S_z \leq S$. 
The individual isospins are coupled to total isospin $T$ and isospin projection $T_z$ in an identical way.
	Next, using the decomposition of a wave vector,
\beq
	| \kvec \ra = 4 \pi \sum_{l m} i^l
	| k l m \ra Y^*_{lm} (\khat) \; ,
\eeq
where $k$ is the magnitude, $l$ is the orbital angular momentum, $m$ is the orbital angular momentum projection, and $Y^*_{lm}$ are spherical harmonics, the basis can then be recoupled to total angular momentum $J$ and projection $J_z$,
\beq
	| l m S S_z \ra
	= \sum_{J J_z}
	\C^{J J_z}_{l m S S_z} | J J_z \ra 
	\; .
\eeq
	Our single-particle states are then expressed as,
\beq
	| 1 2 \ra =
	4 \pi
	\sum_{lSJT}
	\sum_{m S_z}
	\sum_{J_z T_z}
	| k \Pvec (l S) J J_z T T_z \ra
	i^l \;
	Y^*_{lm} (\khat)
\eeq
where we use the short hand $|1 \ra = | \pvec_1 \s_1 \s_{1z} \is_1 \is_{1z} \ra$.
	
	In this basis, we enforce various symmetries of our interaction $\vnn$, namely 
	translational, Galilean, and rotational invariance, conservation of total spin,
	and isospin invariance and charge independence.  After also incorporating the Pauli
	principle, matrix elements of $\vnn$ satisfy
\begin{align*}
	\la k' \Pvec' (l' S') J' J_z' T' T_z'
	| \vnn |
	k \Pvec (l S) J J_z T T_z \ra 
	 & = 
	\la k' (l' S') J' T' 
	| \vnn |
	k (l S) J T \ra
	\left(
	1 - (-1)^{l + S + T}
	\right)
	\\
	&\times 
	(2\pi)^3 \delta^3(\Pvec - \Pvec') 
	\;
	\delta_{S,S'}
	\delta_{J,J'} \delta_{J_z, J_z'}
	\delta_{T,T'} \delta_{T_z,T_z'} \;.
	\numberthis
	\label{eq:pot_pwbasis}
\end{align*}

\subsection{Particle-Particle and hole-hole simplification}

	When angle-averaging the Pauli blocking operators $Q_{\pm}$, the special form of the pp and hh ladders ensures that different partial waves in the ladder do not couple together unless the potential couples them explicitly. 
	This occurs because each single-particle label is uniquely matched with another one in a given bra and ket, i.e., for a given pp or hh diagram in the ladder, any two lines which leave a potential together also enter a potential together.
	To see this, we look at a particular pair of single-particle labels occurring in the interior of the ladder,
\beq
	\sum_{\s \is}
	\int \frac{d^3 \pvec_1}
	{(2\pi)^3} \; 
	\frac{d^3 \pvec_2}
	{{(2\pi)^3}}
	\;
	| 1 2 \ra \la 1 2 | \; Q (p_1, p_2)
\eeq
where $Q$ is either a hole or particle Pauli blocking operator and we have inserted the sums over the single-particle numbers. 
	Going to a partial wave basis and only keeping the relevant quantum numbers in the bra and ket for potential matrix elements we get,
\begin{align*}
	(4\pi)^2
	\sum_{\s \is}
	\int \frac{d^3 \kvec}
	{(2\pi)^3} \; 
	\frac{d^3 \Pvec}
	{{(2\pi)^3}}
	\sum_{S S'}
	\sum_{J J'}
	\sum_{l l'}
	\sum_{m m'}
	\sum_{T T'}
	\sum_{T_z T_z'}
	\sum_{S_z S_z'}
	\sum_{J_z J_z'}
	| k (l S) J T \ra
	\la k (l' S') J' T' |
	\;
	i^{l-l'}
	\\
	\times Q(\Pvec/2, \kvec; \kf) \; Y_{l'm'} (\khat)
	Y_{lm}^* (\khat) \;
	\C_{\s_1 \s_{1z} \s_2 \s_{2z}}^{S S_z}
	\C_{\s_1 \s_{1z} \s_2 \s_{2z}}^{S' S_z'}
	\C_{l m S S_z}^{J J_z}
	\C_{l' m' S' S_z'}^{J' J_z'}
	\C_{\is_1 \is_{1z} \is_2 \is_{2z}}^{T T_z}
	\C_{\is_1 \is_{1z} \is_2 \is_{2z}}^{T' T_z'} \; .
	\numberthis
\end{align*}
	The sums over the single-particle spins and isospins with CG orthogonality fix $T=T'$ and $S=S'$ along with $T_z=T_z'$ and $S_z=S_z'$,
\begin{align*}
	\frac{2}{\pi}
	\int d^3 \kvec
	\frac{d^3 \Pvec}{(2\pi)^3}
	\sum_{S S_z}
	\sum_{J J'}
	\sum_{l l'}
	\sum_{m m'}
	\sum_{T T_z}
	\sum_{J_z J_z'}
	| k (l S) J T \ra
	\la k (l' S) J' T |
	\;
	\\ 
	\times \;
	Q(\Pvec/2, \kvec; \kf) \; 
	i^{l-l'}
	Y_{l'm'} (\khat)
	Y_{lm}^* (\khat) \;
	\C_{l m S S_z}^{J J_z}
	\C_{l' m' S S_z}^{J' J_z'} \; .
	\numberthis
	\label{eq:B13sum}
\end{align*}
	Next we make the assumption that all Pauli blockers are angle-averaged, 
\beq
	Q_{\pm} (\Pvec/2, \kvec; \kf) \to \Qav_{\pm} (P,k;\kf) \; ,
	\label{eq:q_assumption}
\eeq
	such that $d\Omega_\kvec$ dependence is  only present in the spherical harmonics.
	Note that Eq.~\eqref{eq:q_assumption} is automatic if we are considering scattering in free-space ($Q_+ \to 1$) or if our potential is pure s-wave ($l = 0$). 
	With no other angular dependence in the integrand, the solid angle integration can then be done using spherical harmonics orthogonality,
\beq
	\int d\Omega_\kvec  \;
	Y_{l'm'} (\khat)
	Y_{lm}^* (\khat)
	= \delta_{l l'} \; \delta_{m m'} \; .
\eeq
	This simplifies Eq.~\eqref{eq:B13sum} to 
\begin{align*}
	\frac{2}{\pi}
	\int dk \; k^2 \;
	\int \frac{d^3\Pvec}{(2\pi)^3}
	\Qav (P,k;\kf) \; 
	\sum_{S S_z}
	\sum_{J J'}
	\sum_{l m}
	\sum_{T T_z}
	\sum_{J_z J_z'}
	| k (l S) J T \ra
	\la k (l S) J' T |
	\;
	\C_{l m S S_z}^{J J_z}
	\C_{l m S S_z}^{J' J_z'} \; ,
	\numberthis
\end{align*}
and CG orthogonality requires that $J=J'$ along with $J_z = J_z'$.
	Therefore, when considering pp or hh ladder diagrams with the angle-averaging approximation for Pauli blockers, each partial wave not coupled by the potential factorizes,
\begin{align*}
	\frac{2}{\pi}
	\int dk \; k^2 \;
	\sum_{l}
	\Bigg[
	\int \frac{d^3\Pvec}{(2\pi)^3}
	\sum_{S J T}
	\sum_{J_z T_z}
	\Bigg] \;
	| k (l S) J T \ra
	\la k (l S) J T |
	\; \;
	\Qav (P,k;\kf) \; ,
	\numberthis
\end{align*}
	where we have grouped some of the sums in brackets for clarity. 
	The bracketed quantum numbers are diagonal across our $NN$ potential matrix elements and hence only one sum will contribute to any given Goldstone energy diagram.

\section{Hole-Hole channel}

\label{sec:hole_hole}

\begin{figure}[t]
  \includegraphics[width=1.0\textwidth]{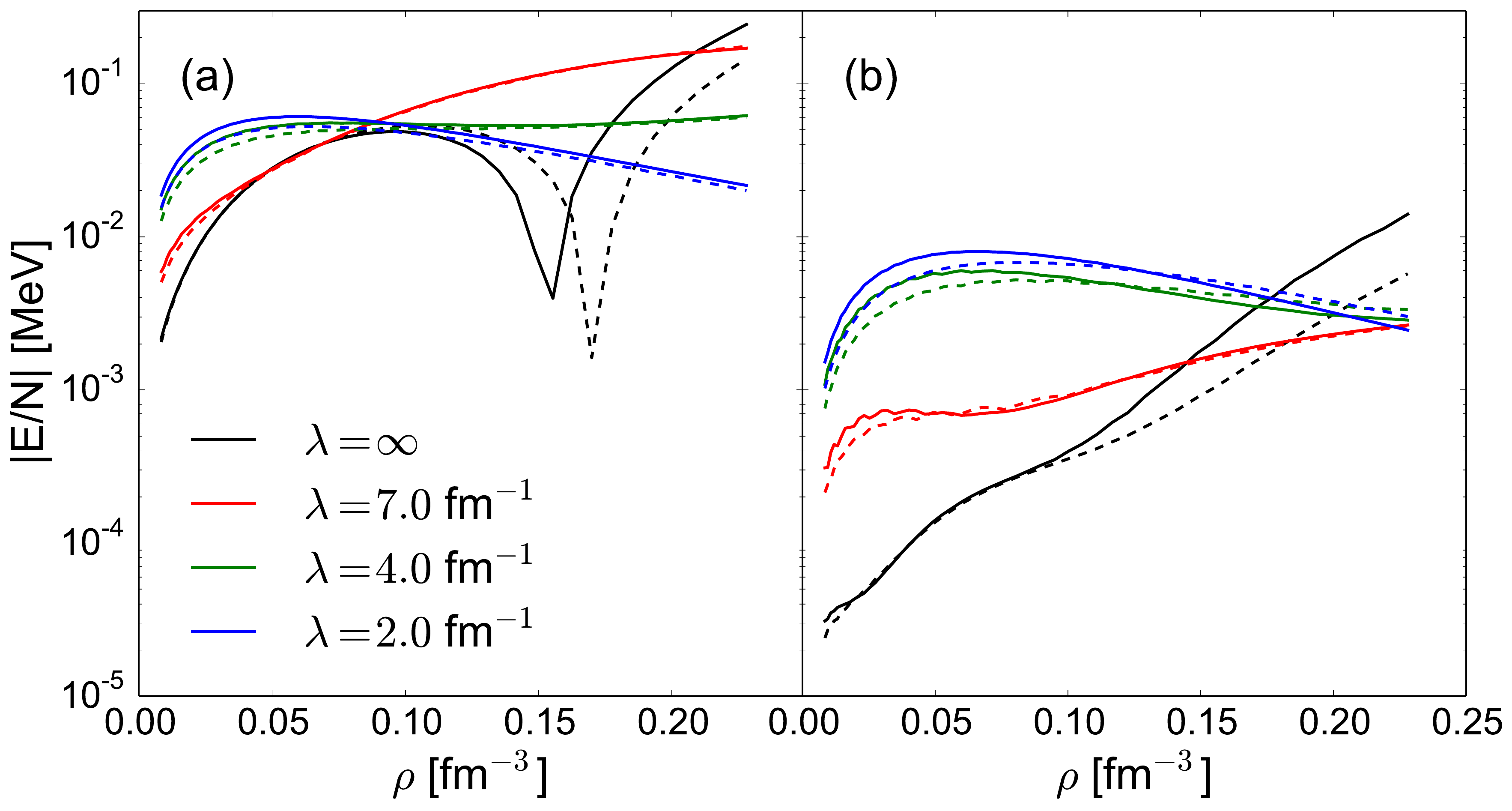}~
  \caption{\textbf{(a)} The absolute value of the third-order energy per particle in nuclear matter for the hh channel is plotted as a function of density $\rho$ for the $\osz$ partial wave using the AV18 potential.  
  Both exact (solid) and estimates (dashed) are shown for four different SRG $\lambda$ scales.
  Estimates are done with $\pmax = 10$.
 \textbf{(b)} Same as (a) but for fourth-order in the hh channel.}
 \label{fig:hh_1S0_AV18}
\end{figure}

\begin{figure}[t]
  \includegraphics[width=1.0\textwidth]{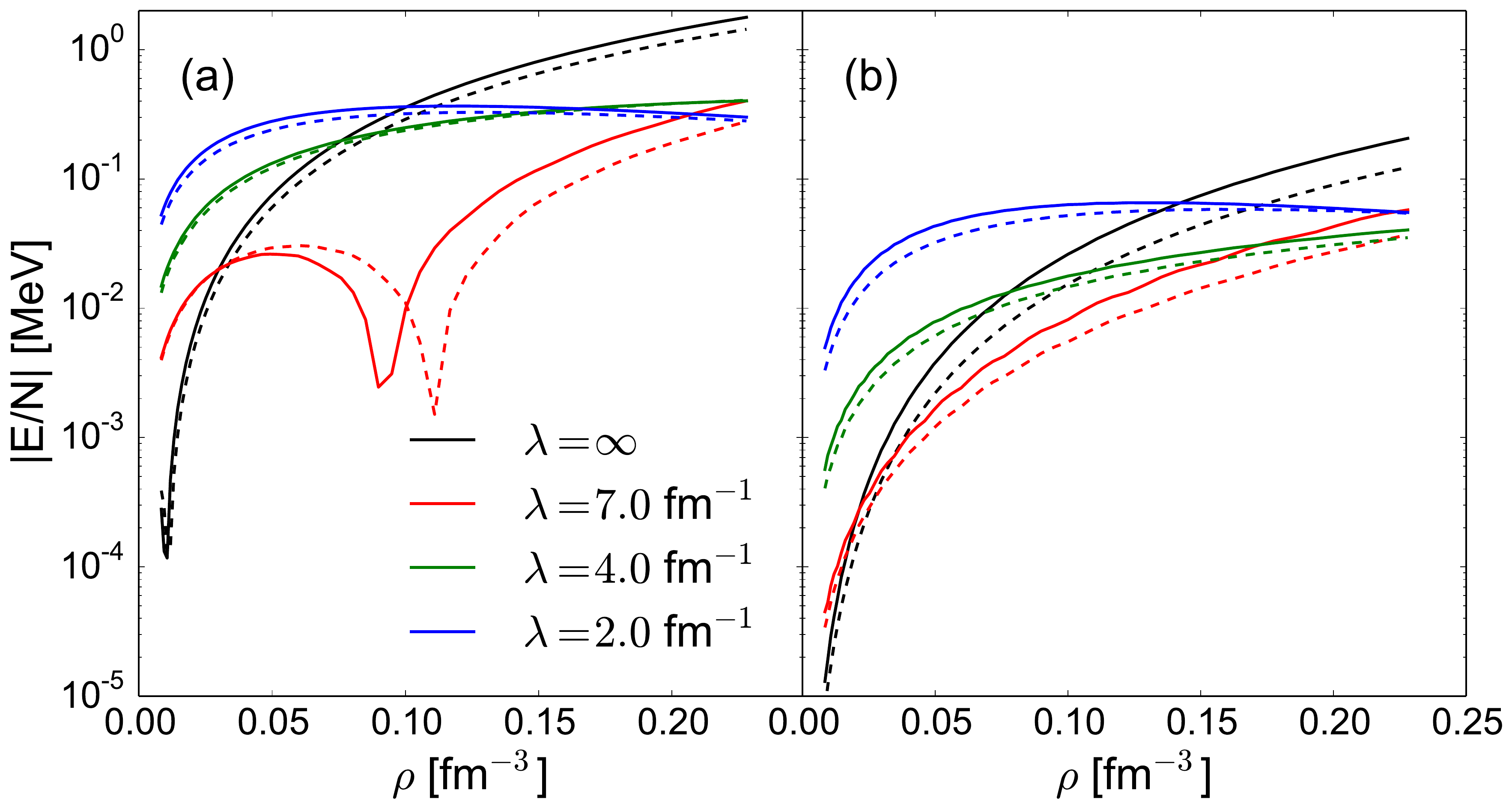}~
  \caption{\textbf{(a)} The absolute value of the third-order energy per particle in nuclear matter for the hh channel is plotted as a function of density $\rho$ for the $\tso$ partial wave using the AV18 potential.  
  Both exact (solid) and estimates (dashed) are shown for four different SRG $\lambda$ scales.
  Estimates are done with $\pmax = 10$.
 \textbf{(b)} Same as (a) but for fourth-order in the hh channel.}
 \label{fig:hh_3S1_AV18}
\end{figure}

\begin{figure}[t]
  \includegraphics[width=0.99\textwidth]{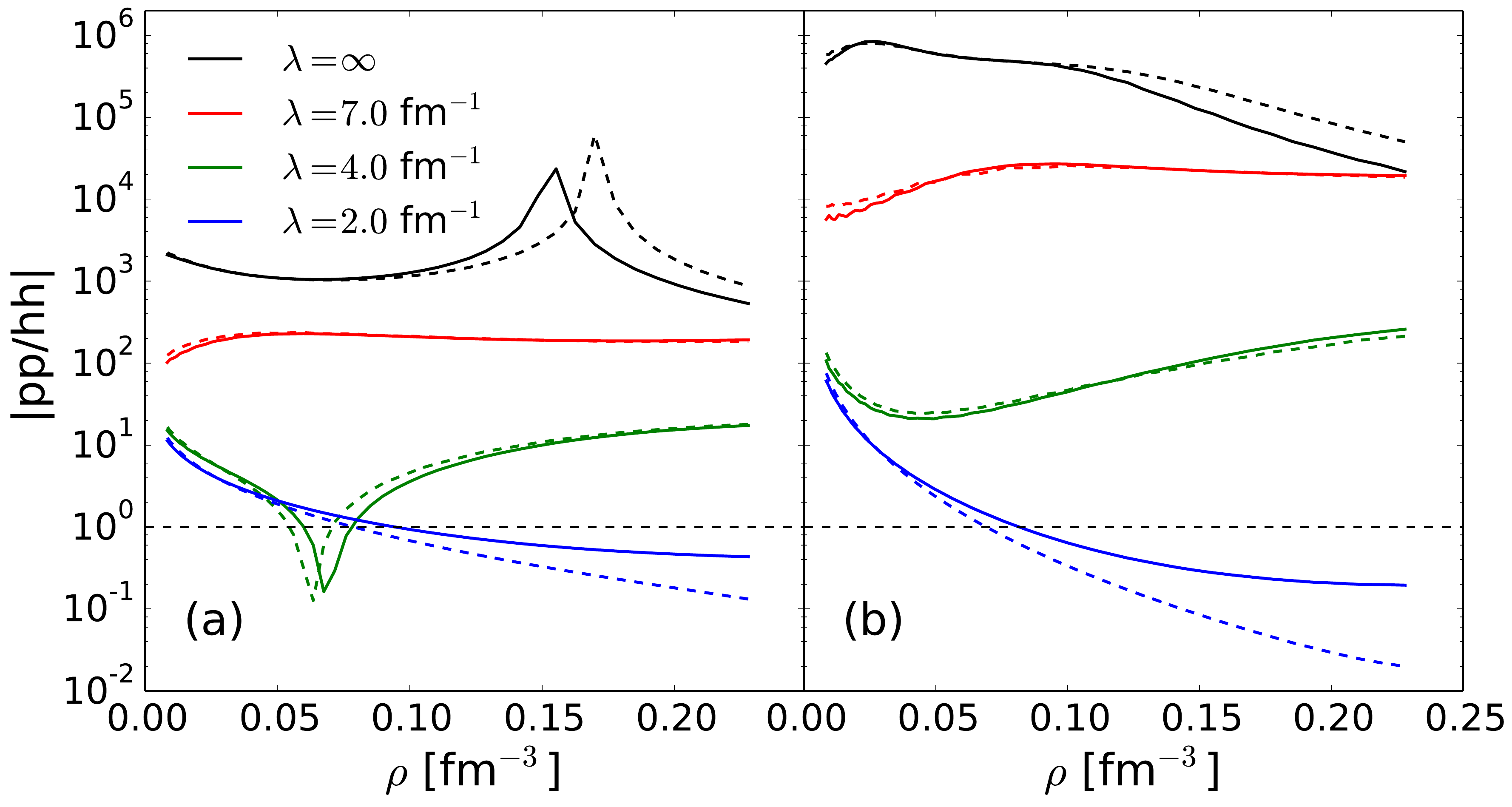}~
  \caption{\textbf{(a)} The absolute value of the ratio of the third-order pp to hh ladder term is plotted as a function of density $\rho$ for the $\osz$ partial wave using the AV18 potential. 
  Both exact (solid) and estimate (dashed) calculations are shown for four different SRG $\lambda$ scales.
  \textbf{(b)} The same as (a) but for fourth-order.
  \label{fig:pphh_1S0}
  }
  
\end{figure}	

\begin{figure}[t]
  \includegraphics[width=0.99\textwidth]{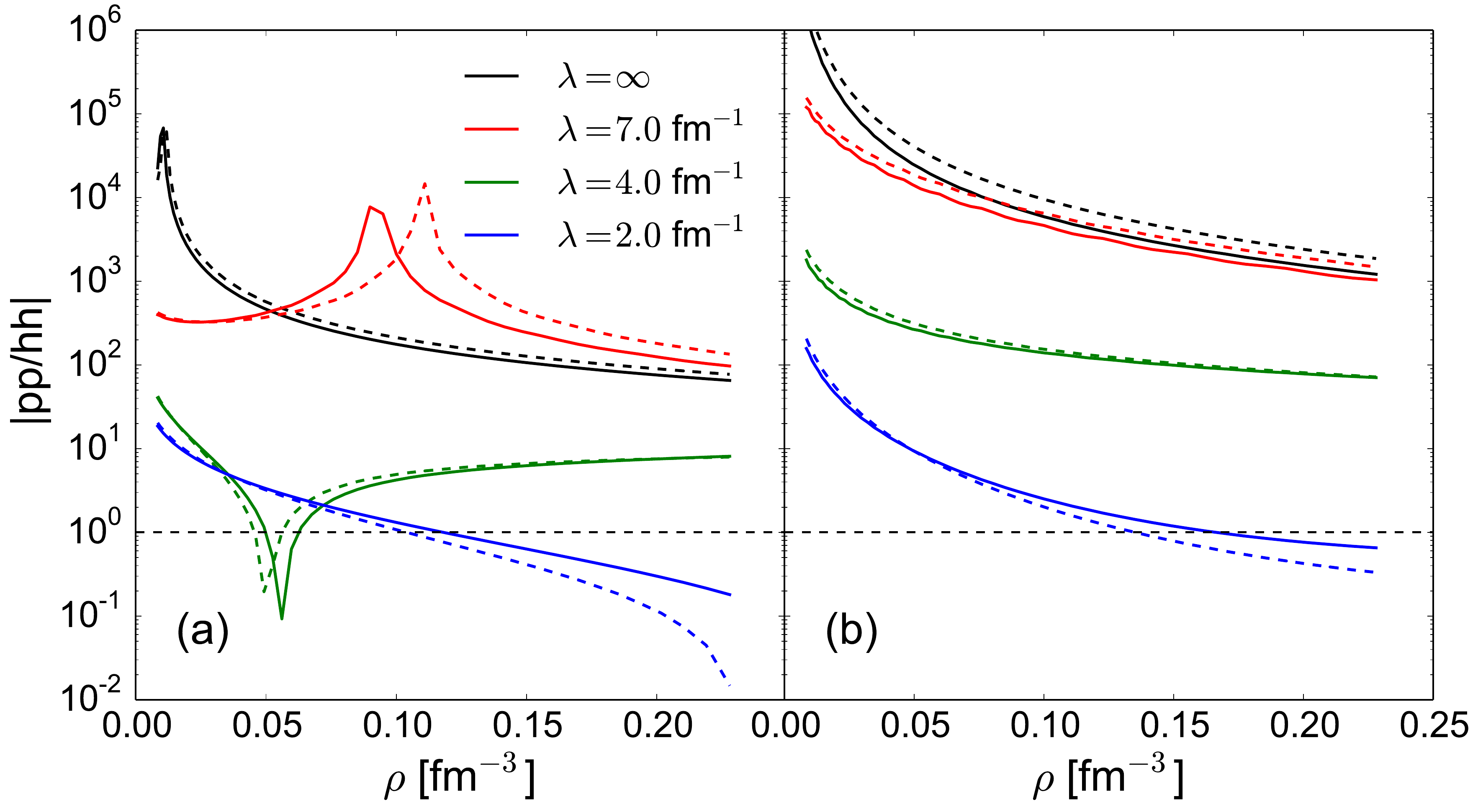}~
  \caption{\textbf{(a)} The absolute value of the ratio of the third-order pp to hh ladder term is plotted as a function of density $\rho$ for the $\tso$ partial wave using the AV18 potential.
  Both exact (solid) and estimate (dashed) calculations are shown for four different SRG $\lambda$ scales.
  \textbf{(b)} The same as (a) but for fourth-order.
  }
  \label{fig:pphh_3S1}
\end{figure}	

The energy per particle of the $n$th rung in the hole-hole ladder assuming angle-averaging for Pauli blockers is given by, 
\begin{align*}
	\frac{E^{(n)}_{\rm hh}}{N}
	&=
	\left(
	\frac{1}{2}
	\right)^{n}
	\left(\frac{2}{\pi}\right)^{n}
	2^{n}
	\left(\frac{m}{\hbar^2} \right)^{n-1}
	\frac{
	\left(
	-1
	\right)^{n-1}
	}{\rho}
	\int 
	\frac{d^3\Pvec}{(2\pi)^3}
	\int dk_1 \; k_1^2 \cdots
	\int dk_{n} \; k_{n}^2 \;
	(2T + 1)
	(2J + 1)
	\\
	&\times
	\ddfrac{\Qavp (P, k_1; \kf) \;
	\Qavh (P, k_2; \kf)
	\cdots
	\Qavh (P, k_{n}; \kf)
	\; 	
	}
	{(k_1^2 - k_2^2) 
	\cdots 
	(k_1^2 - k_n^2)}
	\;
	\la k_1 | V | k_2 \ra 
	\cdots
	\la k_n | V | k_1 \ra
	\; .
	\numberthis
	\label{eq:hh_ladder_energy}
\end{align*}
	To estimate these diagrams, energy denominators and the particle Pauli blocker are approximated as in the pp case,
\beq
	\frac{1}{k_p^2 - k_h^2}
	\approx
	\frac{1}{k_p^2 - \kav^2} \; ,
	\qquad
	\Qavp (P, k; \kf) 
	\approx
	\Qavp (\Pav, k; \kf) \; .
	\label{eq:hh_approximations}
\eeq
	Note that this averaging of momentum magnitudes automatically allows for factorization of the interior hole ladder from the outer particle lines.
	However, handling the multiple hole Pauli blockers $\Qavh (P,k;\kf)$ requires more care. 
	Both the total momentum $P$ and the hole relative momentum $k$ are of order $\kf$, resulting in no obvious factorization for the product of multiple hole Pauli blockers.
	We have not found a way to average $\Qavh$ that consistently reproduces the energy of a given rung in the ladder.
	To estimate the energy diagram of a given rung in the hh ladder, we explicitly keep the total momentum $P$ integral and its dependence in the hole Pauli blockers. 
	All momentum integrals are represented discretely on a Gauss-Legendre mesh with weights $w$.
	The total momentum is summed over the interval 0 to $2\kf$ with total number of points $\pmax$, 
\bseq
\begin{align*}
	&\frac{E^{(n)}_{\rm hh}}{N}
	\approx
	\frac{2}{\pi}
	\left(\frac{2m}{\pi \hbar^2}\right)^{n-1}
	\frac{
	\left(
	-1
	\right)^{n-1}
	}{2 \pi^2 \rho}
	\sum_{k_i}^n
	(2T+1)
	(2J+1)
	\; 
	\Qavp (k_1, \Pav) \;
	k_1^2 \; w_1
	\left(
	\ddfrac{1}{k_1^2 - \kav^2}
	\right)^{n-1}
	\\
	&\times
	\la k_1 | V | k_2 \ra
	\sum_j^{\pmax}
	\left(
	P_j^2 w_j \;
	\sqrt{\Qavh (P_j, k_2; \kf) \; k_2^2 \ w_2}
	\;
	F_j^{n-2}
	\;
	\sqrt{
	\Qavh (P_j, k_n; \kf) \; k_n^2 \ w_n}
	\right)
	\la k_n | V | k_1 \ra
	\; ,
	\numberthis
	\label{eq:hh_approx}
\end{align*}
where the hh kernel $F$ is given by,
\beq
	F_j =
	\sqrt{\Qavh (P_j, k_a; \kf) \; k_a^2 \; w_a} \;
	\la k_a | V | k_b \ra \;
	\sqrt{\Qavh (P_j, k_b; \kf) \; k_b^2 \; w_b} \; .
\eeq
\eseq	
	Each of the individual $j$ pieces in the parentheses is decoupled from each other and can be computed independently. 
	As in the pp case, the hh kernel can be diagonalized as $F$ is real and symmetric,  
\beq
	F^{n}_j =
	L_j D_j^n L_j^{-1} \; ,
\eeq	
	where an additional hh rung for a given value of the total momentum $P_j$ corresponds to an additional power of the eigenvalue matrix $D_j$.
	Once $\pmax$ is set, all rungs of the hh ladder in Eq.~\eqref{eq:hh_approx} carry approximately the same computational load.

	We have found that the product of multiple hole Pauli blockers does not seem to show strong sensitivity to the total momentum $P$. 
	Even for a value of $\pmax=3$, good energy reproductions are found up to fourth order. 	
	Exact values for the third- and fourth-order energy per particle in nuclear matter along with our estimates are plotted in Figs.~\ref{fig:hh_1S0_AV18} and \ref{fig:hh_3S1_AV18} for the $\osz$ and $\tso$ partial waves respectively.
	The energy per particle for both waves shows much less sensitivity at moderate densities to the chosen SRG scale, i.e., less than an order of magnitude near saturation.
	Contrast this with the pp channel in Figs.~\ref{fig:pp_1S0_AV18} and \ref{fig:pp_3S1_AV18} where the energy per particle varies by many orders of magnitude depending on $\lambda$. 
	To compare the relative importance of terms in the two ladders, in Fig.~\ref{fig:pphh_1S0} we plot the absolute value of the ratio of the third-order term in the pp ladder to the third-order term in the hh ladder along with the same for fourth order in the $\osz$ partial wave.
	We plot the same quantities in Fig.~\ref{fig:pphh_3S1} for the $\tso$ partial wave as well. 
	As expected, for the unevolved potential in both partial waves, the pp channel dominates over hh both at third- and fourth-order.
	However as $\lambda$ lowers, the importance of hh ladder terms increases until around $\lambda \approx 2.0 \fmi$ where they are comparable or even larger than the matching pp terms.
	Interestingly, this appears to happen around the same SRG scale for both the $\osz$ and $\tso$ channels again suggesting that phase space is the determining factor.

\begin{figure}[t]
  \includegraphics[width=0.49\textwidth]{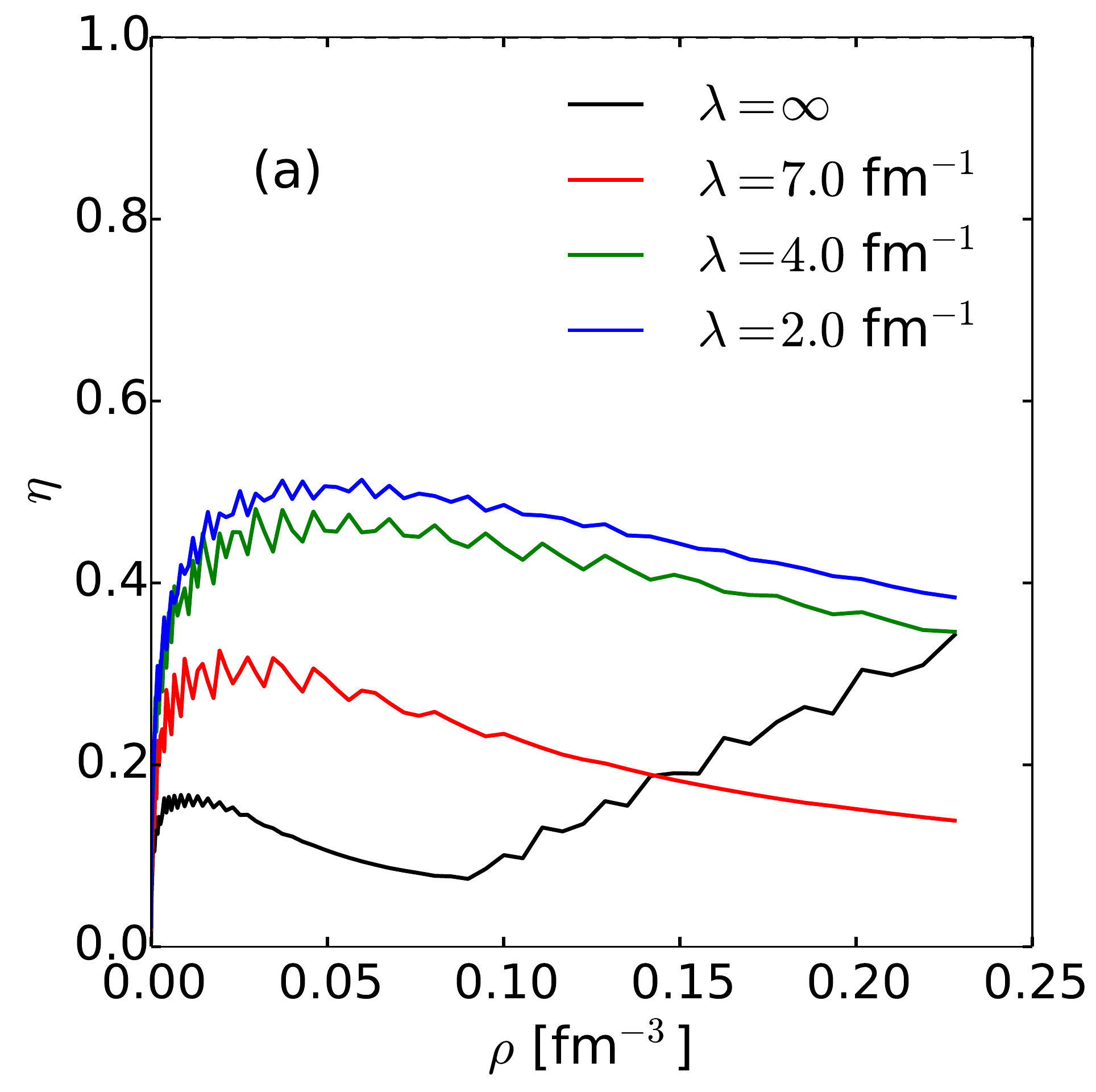}~
  \includegraphics[width=0.49\textwidth]{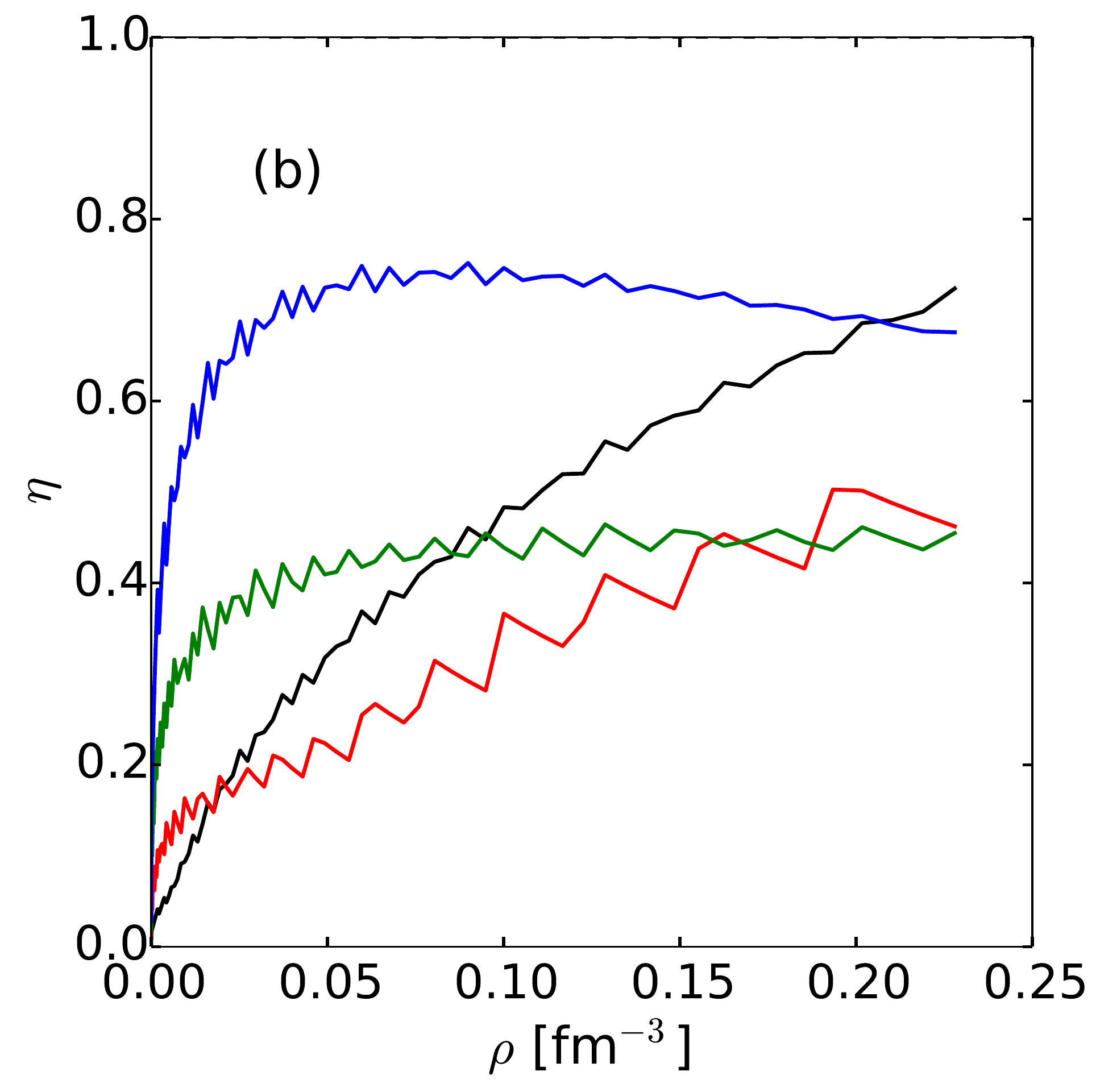}
  \caption{\textbf{(a)} The expansion parameter $\eta$ in Eq.~\eqref{eq:eta_hh} is plotted as a function of density $\rho$ for the $\osz$ partial wave using the AV18 potential. 
  Four different SRG $\lambda$ scales are shown.
  \textbf{(b)} The same as (a) but for the $\tso$ partial wave.
  }
     
  \label{fig:hh_eign_AV18}
\end{figure}

	The hh channel also appears perturbative at all shown densities for each SRG scale just in looking at the relative size of the third- and fourth-order contributions.
	We can attempt to make this statement more rigorous by extracting an expansion parameter from Eq.~\ref{eq:hh_approx} though this is more involved than for the pp case. 
	There are two complications here that are absent in the pp channel:
\be
	\item By keeping the center of momentum $P$ as an explicit variable, each value for $P_j$ will create a different eigenvalue matrix $D_j$. 
	The sum over different $j$ means that the contributions from the different matrices do not factorize when calculating $E^{(n+1)}_{\rm hh} / E^{(n)}_{\rm hh}$. 
	To circumvent this, we look at all the matrices $D_j$ and take the largest eigenvalue $\epsilon_{\text{max}}$ among this set.
	This is motivated by the observation that this maximal value will control behavior in the ladder for high orders.
	\item The outer parts of the integrand in Eq.~\ref{eq:hh_approx} now scale with the number of rungs in the ladder due to the energy denominators, 
	\beq
		\frac{1}{k_1^2 - \kav^2} \; .
		\label{eq:denom_issue}
	\eeq
	Although the outer part of the integrand can be diagonalized similarly to the hh kernel, there will be mixing between the different eigenvalues of the two matrices.
	Instead we create a `worst-case' value for the expansion parameter by approximating the momentum $k_1$ by its smallest value $k_{1,\text{min}}$ allowed by Pauli blocking,
	\beq
		k_{1,\text{min}} = \sqrt{
	\kf^2 - \Pav^2/4	
	}
	\eeq	
such that Eq.~\ref{eq:denom_issue} is maximized. 
\ee

	Our expansion parameter $\eta$ for the hh channel is then given by,
\beq
	\eta = \frac{2m}{\pi \hbar^2}
	\frac{|\epsilon_{\text{max}}|}
	{k_{1,\text{min}}^2 - \kav^2} \; .
	\label{eq:eta_hh}
\eeq
	This value is plotted as a function of density in Fig.~\ref{fig:hh_eign_AV18} for the $\osz$ and $\tso$ partial waves.
	As can be seen, even for our `worst-case' analysis, the hh channel is perturbative for all four SRG scales all the way up to saturation.
	We speculate that this can be primarily attributed to the smaller region of hole-hole phase space available to the system which is unaffected by the running of the SRG.

\section{Angle-averaging \texorpdfstring{$3N$}{3N} Fermi spheres}

	\label{sec:three_body_fermi_spheres}

	In this appendix, we angle-average the $3N$ Pauli blockers.
	We first consider the case of Pauli blocking for three holes and then for three particles. 

	\subsection{Three Holes}

	First we do the integral over the solid angle $\khat$,
\bseq
\begin{align*}
	\frac{1}{4\pi}
	\int d\Omega_\kvec \;
	\Qh \left(
	\Wvec/3 - \jvec/2, \kvec
	; \kf
	\right)
	=
	B 
	\qquad
	\text{with} \qquad
	0 \leq B \leq 1
	\; ,
	\numberthis
\end{align*}
\beq
	B = 
	\frac{\kf^2 - k^2 - W^2/9 - j^2/4 + \Wvec \cdot \jvec/3}
	{k \sqrt{4W^2/9 + j^2 - 4\Wvec \cdot \jvec/3}} \; .
\eeq
\eseq
	Solving for the bounds on $B$ in terms of the angle term $\cos \theta_{Wj}$, hereafter just called $\cos \theta$, gives,
\bseq
\beq
	\cos \theta =  \alpha_3 \; ,
	\qquad
	\cos \theta = R_+ \text{ or } R_- \;,
\eeq
\beq
	\alpha_3 (\kf, W, j, k) \equiv 	
	\dfrac{3}{W j}
	\left( 
	W^2/9 + j^2/4 - \kf^2 + k^2
	\right) \; ,
\eeq
\beq
	R_{\pm} (\kf, W, j, k) \equiv  	
	\dfrac{3}{W j} 
	\left(
	W^2/9 + j^2/4 \pm 2 k \kf - \kf^2 - k^2 
	\right) \; ,
\eeq
\eseq
	where $\alpha_3$ is for $B = 0$ and $R_+$ and $R_-$ are the two roots for $B=1$. 

Now the angular integral over $\jhat$ can be done,
\beq
	\Qth (W, k, j) =
	\frac{1}{4\pi}
	\int d\Omega_\jvec \;
	I_{\text{h}}
	\qquad
	\text{with}
	\qquad
	I_{\text{h}} \equiv 
	n (\Wvec/3 + \jvec)	\times
	B \; ,
\eeq
where $\Qth$ is the fully angle-averaged term and the constraint on $B$ in $I_{\text{h}}$ is implicit.
	The integral over $B$ is given by, substituting $\chi = \cos \theta$,
\beq
	\int B d\chi = 
	\dfrac{
	\sqrt{9 j^2 - 12 \chi j W + 4 W^2}
	\left(
	9 j^2 - 12 \chi j W + 4 
	(
	27 k^2 - 27 \kf^2 + W^2
	)
	\right)	
	}
	{216 W k j}
	\equiv
	\beta (\chi) \; .
\eeq
	From the Fermi sphere term, $n (\Wvec/3 + \jvec)$, we get a constraint on the angle,
\bseq
\beq
	\text{if }
	\cos \theta \leq
	\alpha_1 
	\text{\; then \; $I_{\text{h}}= 1 \times B$ \; otherwise \; $I_{\text{h}}=0$} \; ,
\eeq
\beq
	\alpha_1 (\kf, W, j) \equiv 
	 \dfrac{3}{2 W j}
	 \left(
	\kf^2 - W^2/9 - j^2
	\right) \;  .
\eeq
\eseq
	Observing that $\alpha_3 > R_-$ from the $B$ bounds above, we get the constraints that
\bseq
\beq
	\text{if } \cos \theta < \alpha_3 \text{ then $I_{\text{h}} = 0$} \; , 
\eeq
\beq
	\text{if } \cos \theta > \alpha_1 \text{ then $I_{\text{h}} = 0$} \; , 	
\eeq
\beq	
	\text{if } \cos \theta > R_+ \text{ then $I_{\text{h}} = 1$} \; , 
\eeq
\beq
	\text{otherwise } I_{\text{h}} = B \; .
\eeq
\eseq
	Note that keeping $B$ non-negative and avoiding the branch point in the denominator requires that $k < \kf$ which in turn implies that $R_+ > \alpha_3$.
	The two different possible orderings of the constraints are then:
$\alpha_3 < R_+ < \alpha_1$ and 
$\alpha_3 < \alpha_1 < R_+$. 
	From these constraints, the piecewise values of $\Qth$ are then,
\bseq
\beq
	\Qth = 	
	\;
	\frac{1}{2}
	\begin{cases}
	0 &  \text{if \; $\alpha_3 > 1$ \; or \; $\alpha_1 < -1$ \; or \; $\alpha_3 > \alpha_1$} \; ,
	\\
	t_u + 1
	& \text{if $R_+ < \alpha_1$ and $R_+ < -1$}
	\\
	\beta(1) - 
	\beta(t_l)
	& \text{if $R_+ < \alpha_1$ and $R_+ > 1$}
	\\
	\beta (t_u)
	-
	\beta (t_l)
	& \text{if $\alpha_1 < R_+$}
	\\
	\beta(R_+) -
	\beta(t_l) + t_u 
	- R_+
	& \text{otherwise}
	\end{cases}
	\; ,
	\numberthis
\eeq
\beq
	t_u = \text{min}	
	\{\alpha_1, 1 \}
	\; ,
	\qquad  
	t_l = \text{max}
	\{\alpha_3,-1 \} \; .
\eeq
\eseq

\subsection{Three Particles}

	Again, we first do the solid angle integral over $\khat$,
\begin{align*}
	\frac{1}{4\pi}
	\int d\Omega_{\kvec} \;
	\Qp \left(
	\Wvec/3 - \jvec/2, \kvec
	; \kf
	\right) \;
	=
	\;
	\overline{B}
	\quad
	\text{with}
	\quad
	0 \leq \bbar \leq 1
	\quad
	\text{where}
	\quad
	\bbar \equiv -B \; ,
	\numberthis
\end{align*}
and then the solid angle integral $\jhat$,
\beq
	\Qtp (W, k, j) =
	\frac{1}{4\pi}
	\int d\Omega_{\jvec} \;
	I_{\text{p}}
	\qquad
	\text{where}
	\qquad
	I_{\text{p}} \equiv 
	\nb (\Wvec/3 + \jvec)	\times
	\bbar \; ,
\eeq
	where now the constraint from the lone Fermi sphere flips the sign from the hh case,
\beq
	\text{if}
	\quad
	\cos \theta \geq \alpha_1 (\kf, W, j)
	\quad
	\text{\; then \; $I_{\text{p}} = 1 \times \bbar$ \; otherwise \; $I_{\text{p}}=0$} \; .
\eeq

	Now, note that the sign flip in $\bbar$ implies two different sets of constraints depending on the magnitudes of $k$ and $\kf$ due to the branch point.
	For $k < \kf$, and looking at the bounds on $B$ found above,
\bseq
\beq
	\text{if } \cos \theta < \alpha_1 \text{ then $I_{\text{p}} = 0$} \; , 
\eeq
\beq
	\text{if } \cos \theta > \alpha_3 \text{ then $I_{\text{p}} = 0$} \; , 	
\eeq
\beq	
	\text{if } \cos \theta < R_- \text{ then $I_{\text{p}} = 1$} \; , 
\eeq
\beq
	\text{otherwise } I_{\text{p}} = \bbar \; ,
\eeq
\eseq
	and gives the orderings $\alpha_1 < R_- < \alpha_3$ and $R_- < \alpha_1 < \alpha_3$. 
	We also get the constraints for $k > \kf$,
\bseq
\beq
	\text{if } \cos \theta < \alpha_1 \text{ then $I_{\text{p}} = 0$} \; , 
\eeq
\beq
	\text{if } \cos \theta < R_- \text{ then $I_{\text{p}} = 1$} \; , 	
\eeq
\beq	
	\text{if } \cos \theta > R_+ \text{ then $I_{\text{p}} = 1$} \; , 
\eeq
\beq
	\text{otherwise } I_{\text{p}} = \bbar \; .
\eeq
\eseq	
	giving the orderings $\alpha_1 < R_- < R_+$, and $R_- < \alpha_1 < R_+$, and $R_- < R_+ < \alpha_1$.
	Using the following notations,
\bseq
\beq
	\Qtp = \Qtp^1 + \Qtp^2
\eeq
\beq	
	\bebar (\chi) \equiv - \beta (\chi) \; ,
	\quad
	\Qtp^1 \equiv
	\frac{1}{4\pi}
	\int d\Omega_\jvec \;
	I_{\text{p}} \; \theta(\kf - k) \; ,
	\quad
	\Qtp^2 \equiv
	\frac{1}{4\pi}
	\int d\Omega_\jvec \;
	I_{\text{p}} \; \theta(k - \kf)
\eeq
\beq
	\Qtp^2 = \Qtp^{2,a} + \Qtp^{2,b} \; ,
\eeq
\beq
	\Qtp^{2,a} = \Qtp^2 \; \theta (R_- - \alpha_1) \; ,
	\qquad
	\Qtp^{2,b} = \Qtp^2 \; \theta (\alpha_1 - R_-) \; ,
\eeq
\eseq
	and working with the different possible orderings of the constraints, the angle-averaged terms are given by,
\bseq
\beq
	\Qtp^1 
	=
	\;
	\frac{1}{2}
	\begin{cases}
	0 &
	\text{if $\alpha_1 > 1$ or $\alpha_3 < -1$ or $\alpha_1 > \alpha_3$} 
	\\
	1 - p_l &
	\text{if $R_- >  \alpha_1$ and $R_- > 1$}
	\\
	\bebar (p_u) - \bebar (-1) &
	\text{if $R_- >  \alpha_1$ and $R_- < -1$}
	\\
	\bebar(p_u) - \bebar (p_l) &
	\text{if $\alpha_1 > R_-$}
	\\
	\bebar (p_u) - \bebar (R_-)
	+ R_{-} - p_l
	& \text{otherwise} \;
	\end{cases}
	\; ,
\eeq
\beq
	\Qtp^{2,a} = 	\;
	\frac{1}{2}
	\begin{cases}
	0 &
	\text{if $\alpha_1 > 1$} 
	\\
	1 - p_l &
	\text{if $R_- > 1$}
	\\
	\bebar (1) - \bebar (-1) &
	\text{if $R_- < -1$ and $R_+ > 1$}
	\\
	2 &
	\text{if $R_+ < -1$}
	\\
	\bebar(1) - \bebar(R_-) + R_- - p_l  &
	\text{if $R_- <  1$ and $R_+ > 1$}
	\\
	\bebar(R_+) - \bebar(-1) + 1 - R_+ &
	\text{if $R_- <  -1$ and $R_+ < 1$}
	\\
	\bebar (R_+) - \bebar (R_-)
	+ 1 - R_+ + R_{-} - p_l
	& \text{otherwise} \;
	\end{cases}
	\; ,
\eeq
\beq
	\Qtp^{2,b} = 	\;
	\frac{1}{2}
	\begin{cases}
	0 &
	\text{if $\alpha_1 > 1$} 
	\\
	1 - p_l &
	\text{if $\alpha_1 > R_+$}
	\\
	\bebar (1) - \bebar (p_l) &
	\text{if $R_+ > \alpha_1$ and $R_+ > 1$}
	\\
	2 &
	\text{if $R_+ > \alpha_1$ and $R_+ < -1$}
	\\
	\bebar (R_+) - \bebar (p_l)
	+ 1 - R_+ 
	& \text{otherwise} \;
	\end{cases}
	\; ,
\eeq
\beq
	p_u = \text{min}	
	\{\alpha_3, 1 \}
	\; ,
	\qquad  
	p_l = \text{max}
	\{\alpha_1,-1 \} \; .
\eeq
\eseq

\end{document}